 \documentclass[final,5p,times,twocolumn,authoryear]{elsarticle}

\usepackage{newtxtext,newtxmath}
\usepackage{lipsum}
\usepackage{graphicx}	
\usepackage{amsmath}	
\usepackage[normalem]{ulem}
\usepackage[T1]{fontenc}
\usepackage{hyperref}

\DeclareRobustCommand{\VAN}[3]{#2}
\let\VANthebibliography\thebibliography
\def\thebibliography{\DeclareRobustCommand{\VAN}[3]{##3}\VANthebibliography}

\newcommand{\be}{\begin{equation}}
\newcommand{\ee}{\end{equation}}
\newcommand{\bary}{\begin{eqnarray}}
\newcommand{\eary}{\end{eqnarray}}

\journal{High Energy Astrophysics}

\begin{document}

\begin{frontmatter}



\title{The multiwavelength correlations quest for central engines of GRB plateaus: magnetar vs black hole spin-down}


\author[first,*]{Aleksander \L{}. Lenart \thanks{abc} }
\author[2nd,3rd,4th,5th,6th]{Maria G. Dainotti}
\author[7th]{Nikita Khatiya}
\author[7th]{Dhruv Bal}
\author[7th]{Dieter H. Hartmann}
\author[8th]{Nissim Fraija}
\author[5th]{Bing Zhang}
\affiliation[first]{organization={Astronomical Observatory, Jagiellonian University},
            addressline={Orla 171}, 
            city={Kraków},
            postcode={30-244}, 
            state={Małopolska},
            country={Poland}}
\affiliation[2nd]{organization={Division of Science, National Astronomical Observatory of Japan},
            addressline={2-21-1 Osawa}, 
            city={Mitaka},
            postcode={181-8588}, 
            state={Tokyo},
            country={Japan}}
\affiliation[3rd]{organization={The Graduate University for Advanced Studies (SOKENDAI)},
            addressline={Shonankokusaimura, Hayama}, 
            city={Kanagawa},
            postcode={240-0115}, 
            state={Miura District},
            country={Japan}}
\affiliation[4th]{organization={Space Science Institute},
            addressline={4765 Walnut St, Suite B}, 
            city={Boulder},
            postcode={CO 80301}, 
            state={Nevada},
            country={USA}}
\affiliation[5th]{organization={Center for Astrophysics, University of Nevada},
            addressline={4505 Maryland Parkway}, 
            city={Las Vegas},
            postcode={NV 89154}, 
            state={Nevada},
            country={USA}}
\affiliation[6th]{organization={Bay Environmental Institute},
            addressline={P.O. Box 25}, 
            city={Moffett Field},
            postcode={CA 94035}, 
            state={California},
            country={USA}}
\affiliation[7th]{organization={Department of Physics \& Astronomy, Clemson University, Kinard Lab of Physics},
            addressline={Delta Epsilon Ct}, 
            city={Clemson},
            postcode={SC 29634}, 
            state={South Carolina},
            country={USA}}
\affiliation[8th]{organization={Instituto de Astronomía, Universidad Nacional Autónoma de México},
            addressline={Circuito Exterior, C.U.}, 
            city={Cd. de México},
            postcode={A. Postal 70-264, 04510}, 
            state={México},
            country={México}}
\affiliation[*]{Corresponding author: aleksander.lenart@student.uj.edu.pl;  lenart.a.lenart@gmail.com}

\begin{abstract}
This manuscript presents a multilevel analysis of gamma-ray bursts (GRBs). We focus on the plateau phase, which is often observed in the light curves (LCs) of GRBs. We discuss its observational properties and then thoroughly examine possible theoretical models to explain them. Inspired by the limitations of many currently known models, we introduce a novel scenario of an LC powered by the kinetic energy of a rotating black hole (BH). We investigate observational correlations between the properties of GRBs across the gamma, X-ray, and optical bands during the prompt and plateau phases of their LCs. Our analysis includes all GRBs with known redshifts detected by the Neil Gehrels {\it Swift} Observatory ({\it Swift}) and the {\it Fermi} Gamma-ray Space Telescope ({\it Fermi}), as well as ground-based optical telescopes. We identify a tight correlation with the $R^2$ coefficient of $\sim 0.89$ for the three-dimensional Dainotti relation between the luminosity at the end of the plateau, its duration measured by {\it Swift}, and the peak luminosity measured by {\it Fermi} in the 10-1000 keV band. When accounting for redshift evolution, we achieve very small intrinsic scatter $\sigma_{int}=0.25\pm0.04$ ($\sim 43\%$ reduction compared to the previous results). Additionally, we explore correlations involving the optical luminosity at the end of the plateau, yielding promising results. We investigate the clustering of different classes of GRBs in the investigated parameter space and discuss its impact on the aforementioned correlations as well as $E_{iso}$-$E^*_{peak}$ correlation. Notably, we demonstrate how to use the correlations as a powerful class discriminator. Finally, we discuss the theory supporting the evidence of the plateau emission. We present a new paradigm for the GRB plateau: energy extraction from a quickly rotating black hole (BH) via spin-down by a magnetically arrested disk (MAD). We compare this model with observations and explain multiple observed features. We predict the plateau luminosity - time anti-correlation and discuss the cosmological evolution within this proposed model. Furthermore, within this new model, we discuss the possible physical origin of the clustering of long and short GRBs in the parameter space of plateau luminosity - time - prompt luminosity. 
\end{abstract}




\end{frontmatter}




\section{Introduction}

Gamma-ray bursts (GRBs) are among the most energetic events in the Universe, first detected in the late 1960s by the Vela satellites. 
The identification of the first optical afterglow for GRB 970228 in 1997 confirmed the cosmological origin of GRBs, showing that these bursts were extragalactic and occurred at vast distances \citep{vanParadijs1997}. This discovery marked a new era in GRB research, allowing for multi-wavelength observations and providing insights into the environments and progenitors of GRBs. Subsequent studies with various observatories, including the launch of the Neil Gehrels {\it Swift} observatory ({\it Swift}) in 2004 \citep{Gehrels2004}, revealed new patterns in the GRB X-ray light curves, such as the ``plateau phase," where the luminosity of the afterglow remains nearly constant for a period before decaying more rapidly \citep{Zhang2006,Nousek2006,Evans2007,OBrien2006,Willingale2007,2007ApJ...666.1002Z,Liang2007, Dainotti2008,Evans2009}.

The plateau phase was first detected in X-rays, but its existence in the optical band was confirmed shortly thereafter {\citep{Vestrand2006} The first detection of an optical plateau was reported for GRB 050801, whose light curve showed a distinct flattening similar to that seen in X-ray observations, suggesting a complex interplay between the central engine of the GRB and its surrounding environment \citep{Rykoff2006,Vestrand2006}. This phase provided key insights into the energy dissipation mechanisms at work in GRBs. It hinted at a potential link between the plateau and the underlying physical processes of the burst. 

GRBs are not only sources observed in multi-wavelengths, but they also offer the unique opportunity to open a window to the early Universe. Indeed, with current instruments, GRBs are observable up to redshift $z\approx9.4$ \cite{Cucchiara2011}. This makes GRBs a promising tool for exploring the early universe, given that we can predict their absolute luminosity with empirical correlations. The \cite{amati2006} or \cite{Ghirlanda2016A&A...594A..84G} relations predict the total amount of energy emitted by the GRB in the prompt phase $E_{\rm iso}$ based on the shape of the spectra. Specifically, there exists a nonlinear correlation between $E_{\rm iso}$ and $E_{\rm peak}$ - the photon energy corresponding to the maximum fluence density $\nu f_{\nu}$. 
\\\cite{Yonetoku2004} presented a similar correlation between $E_{\rm peak}$ and the peak luminosity. 
An independent correlation exists between the X-ray luminosity at the end of the plateau phase $L_X$ and the rest-frame duration of the plateau $T^{*}_{X}$. First, it was discovered by \cite{Dainotti2008} with a sample of 32 GRBs and confirmed with a sample of 77 GRBs by \cite{dainotti2010}. \cite{Dainotti2011} provided a careful study of the possible systematics. \cite{Dainotti2013,Dainotti2015} proved that this correlation exists even when selection and evolutionary effects are accounted for in the variables pertinent to the relations. This two-dimensional correlation was further extended with the addition of a high-energy 1s peak luminosity of the prompt phase $L_{\rm peak}$, establishing the so-called Dainotti relation \citep{dainotti2015a,Dainotti2016}. The most recent and comprehensive study of the Dainotti relation was presented in \cite{Dainotti2020}, where the authors confirmed the existence of the three-dimensional correlation for different classes of GRBs and the high-quality sub-samples of the whole population}. 
This correlation was independently studied in many variants in the literature by \cite{Zaninoni2011,Bernardini2012,Xu2012A&A...538A.134X,Mangano2012,Sultana_2012,Zaninoni2013,Margutti2013,Bardho2015,Izzo2015A&A...582A.115I,Kawakubo_2015,Si_2018,Zhao2019ApJ...883...97Z,Tang2019ApJS..245....1T,WangFeifei2020,Muccino2020,CaoShulei2022a,XuFan2021ApJ...920..135X,Levine2022,Shuang2022ApJ...924...69Y,Deng2023ApJ...943..126D,Tian_2023,Li_2023,Xu_2023,Deng2025}. 
Moreover, either the Dainotti correlation or its variants were already successfully applied to infer the values of cosmological parameters \citep{Cardone2009,Postnikov2014,Zitouni2016,Luongo2020,Cao2021,Luongo2021a,Luongo2021,Hu2021,Khadka2021,Muccino_2021,XuFan2021ApJ...920..135X,Wang2022ApJ...924...97W,CaoShulei2022,Dainotti2023c,Tian_2023,Li_2023,Xu_2023,
2023MNRAS.521.3909B,2023ApJ...951...63D,2024JCAP...08..015A,2024JHEAp..44..323F,LiJia2024,Alfano2024,SUDHARANI2024101522,Sethi2024}.
We stress here that the existence of the plateau phase holds the promise of more precise standardization of these events. 
A separate analysis of the GRB's prompt phase correlations and the afterglow correlations presents us with a unique opportunity to cross-check the standardization of the same sources. It is crucial to pinpoint here that both phases arise due to different physical mechanisms; thus, we can independently test the relation of observable parameters with redshift. We highlight this point since it has been widely demonstrated in the GRB literature that their luminosity, among different properties, does evolve with redshift. The initial study of the evolution of prompt variables has been discussed first by \citep{Lloyd1999,2000ApJ...534..227L,LloydRonning2002}, while for plateau variables it was first studied by \cite{Dainotti2013b,Dainotti2015,Dainotti2017a} who presented relevant comprehensive observational evidence, and \cite{Volpato2024} and \cite{2023ApJ...947...85L} discussed the theoretical interpretation.
It is important to pinpoint how evolution affects the cosmological parameters. In \cite{Dainotti2023c,Lenart2023} and \cite{Bargiacchi2023}, some of us explored how evolution should be included in the treatment to infer cosmological parameters. In the first two mentioned works, the evolution is treated as a function of cosmological parameters, thus enabling their circularity-free estimation. While in the \cite{Bargiacchi2023} the evolution is fixed to a given function.
\cite{Lenart2023} showed with a sample of $\sim$2400 quasars that the existence of such evolution strongly alters the correlation-based cosmological computations. 
Given evolution's degeneracy and the assumed cosmological model, it is vital to test probes that show different rates of evolution. Thus, GRBs as independent probes, with multiple different correlations corrected for evolution with a selection bias-free method, might be the ultimate tool to compare the cosmological results obtained with different probes. However, if GRBs and QSOs show the same behaviour of the cosmological expansion obtained for multiple correlations, we could have found a trace of new physics beyond the flat $\Lambda$CDM. 
This background shows how crucial further developments of both known and new correlations are. 
Indeed, the application of GRBs in cosmology has been a subject of scientific discussion since their discovery in 2004. One of the first measurements of cosmological parameters was performed by \cite{Schaefer2007} with the application of multiple correlations. The very first distance ladder-independent confirmation of dark energy with GRBs was presented by \cite{Amati2008}. Further, the $E_{iso}$-$E^*_{peak}$ correlation was the subject of many cosmological studies as a stand-alone probe with and without calibration. Other measurements have been discussed by \cite{Amati2013,Wang2016,Demianski2017a,Demianski2017b,Cao2024} and are also related to this correlation. The Dainotti correlation was applied as a cosmological tool for the first time by \cite{Cardone2009,Cardone2010,Dainotti2013,Postnikov2014}. Similarly to the $E_{iso}$-$E^*_{peak}$ correlation, the Dainotti 3D relation was used to investigate cosmological models in many variants by many authors, including \cite{Cao2021,CaoShulei2022a, CaoShulei2022} and \cite{Dainotti2023b,Dainotti2022PASJ}. In the last one, a binning analysis was performed. A historical review on the application of GRB relation as a cosmological tool can be found in \cite{Dainotti2017a,Dainotti2018,DainottiAmati2018}. A recent update following these reviews, which also entails the inclusion of the 3D relations, can be found in \cite{Bargiacchi2025}.

In this article, we present a new relation between the X-ray luminosity at the end of the plateau phase ($L_{X}$), the time at the end of the X-ray plateau ($T^{*}_{X}$), and the corresponding optical luminosity ($L_{opt}$). This correlation sheds light on the physics governing GRB emissions and could enhance their use as standard candles for cosmological studies. Moreover, we analyse the sample of GRBs detected by both {\it Swift} and the {\it Fermi} Gamma-ray Space Telescope ({\it Fermi}) \citep{2009ApJ...702..791M,2009ApJ...697.1071A}. We show that the employment of peak luminosity measured by {\it Fermi} ($L_{\text{peak, Fermi}}$) leads to a decrease in the scatter of the Dainotti relation when compared to the correlation obtained for {\it Swift} measurements involving corresponding luminosity ($L_{\text{peak, Swift}}$). We discuss the clustering of sources in the parameter space of the Dainotti and Amati correlations and pinpoint regions occupied by the GRBs of different origins. Traditionally, GRBs are divided into long and short (LGRBs and SGRBs), based on the time during which 90\% of the burst's energy is emitted $T_{90}$ (measured in the source's rest frame). The distribution of this parameter measured by the BATSE instrument has a bimodal shape with a boundary between the two Gaussian-like distributions at $T_{90}\sim 2\,\rm s$. It is considered that LGRBs ($T_{90}> 2\,\rm s$) are a result of the core collapse of massive stars (the so-called collapsars) \citep{mazets1981catalog,kouveliotou1993identification,paczynski1998ApJ...494L..45P, Woosley2006ARA}, while short ($T_{90}< 2\,\rm s$) arise from the merger of two neutron stars (NS) or an NS and a black hole (BH) \citep{1992ApJ...395L..83N, 1992ApJ...392L...9D, 1992Natur.357..472U, 1994MNRAS.270..480T, 2007PhR...442..166N, 2017ApJ...846L...5G, Goldstein2017, 2017ApJ...848L..12A}.
A newer estimate based on the rest frame duration $T^{*}_{90}=T_{90}/(1+z)$ found by \cite{Gomboc2010} based on the {\it Swift} data is $T^{*}_{90}=0.65\,\rm s$. A later estimate found by \cite{Bromberg2013ApJ...764..179B} for {\it Swift} is $T_{90}=0.8\,\rm s$. However, some researchers still use a $2\,\rm s$ boundary. It is clear that such a short boundary does not capture all merger-type GRBs, given the observations of the kilonova associated with GRB with $T^*_{90}>2\,\rm s$ \citep{Zhang2025}.
We found that SGRBs are outliers of both Dainotti and Amati correlations. The two correlations can be applied to pinpoint the origin (merger or collapsar) of a given source. Given that some SGRBs are pinpointed only by one of the two, those correlations complement each other. 
Commonly, it is considered that both scenarios of GRB origin lead to the rise of a BH. Indeed, \cite{Kuroda2018} proved with 3D magnetohydrodynamical simulations that a core collapse of 70 $M_{\odot}$ star produces a black hole. However, as we discuss further in the text, there is multiple evidence in the literature that SGRB's plateau is driven by a magnetar central engine \citep{Troja2007,Lyons2010,Rowlinson2010,Rowlinson2013}. The launch of magnetically dominated jets in core collapse event was also studied via simulations by \cite{Takiwaki2009}, where authors proved that a magnetar is a likely output of a studied physical setup. In the magnetar scenario, the kinetic energy of the engine is extracted via multipole radiation of charge accumulated on the NS surface. The first theoretical prediction of GRBs driven by this process was presented by \cite{DaiLu1998A&A} and \cite{DaiLu1998PhRvL}, while \cite{zhangmeszaros2001} was among the first to predict the plateau phase generated by this mechanism. Further, this model was studied for the first observations by \cite{Zhang2006}, and confronted with LGRBs and SGRBs data by \cite{LuZhang2014,Lu2015}. Remarkably, \cite{Rowlinson2014} presented a comprehensive explanation of the $L_{X}$-$T^*_{X}$ anti-correlation within the magnetar spin-down model. 
\cite{Ruffini2014} also successfully explained this observational correlation for some of the GRBs within the so-called binary-driven hypernova model. \cite{DereliBegue2022} discuss that the plateau originates from emission that occurs during the coasting phase of the propagating forward shock. This phase precedes the self-similar decaying phase that produces the late-time afterglow. Its key difference from the classical GRB "fireball" evolution is the assumption that during the coasting phase, the GRB Lorentz factor does not exceed a few tens (rather than the common assumption of a Lorentz factor of a few hundred). Furthermore, this model requires that the explosion occurs in a low-density stellar wind environment, as is indeed expected for massive star GRB progenitors. Therefore, this model effectively explains the plateau phase as being due to synchrotron radiation from particles accelerated by the forward shock during the coasting phase by accounting for both the X-ray and optical signals.
A recent alternative explanation of the luminosity - time correlation was obtained by \cite{Duffell_2020} for the jet interacting with the circumstellar medium, as well as the jet seen off-axis \citep{Beniamini2020}. However, as shown by \cite{Li2018b}, some GRBs are too energetic to be explained by a magnetar model, and thus, they have to be powered by a BH. Moreover, those BH-driven GRBs still tightly follow the $L_{X}$-$T^*_{X}$ relation. Although the interpretation of the plateau via accretion model was presented by \cite{Kumar2008,Cannizzo2009,Cannizzo2011}, the origin of the aforementioned relation remains still to be investigated, given that its existence would demand a constant amount of matter being accreted onto the BH \citep{Cannizzo2011}. For the prompt phase, there are two common scenarios of mechanism that would power the burst: the \cite{BlandfordZnajek} (BZ) and the neutrino annihilation \citep{Popham1999} process. However, \cite{Liu2017} concluded that the neutrino annihilation model is a possible explanation for most SGRBs, only half of the LGRBs, and none of the ultra-long GRBs. An alternative scenario of electron-positron outflow from BH was proposed by \cite{vanPutten2000}.
For the first time, the BZ mechanism was proposed to drive the GRB emission by \cite{LEE200083}, but only in the context of the prompt phase. Later, the community tested this hypothesis via simulations. \cite{Nagataki2009,Nagataki2011} were among the first works that discussed the BZ-driven LGRBs. The author proved with 2D simulations that a quickly rotating BH launches jets energetic enough to explain the observed luminosity of GRBs. However, the presented analysis focuses on the first 10-100s after the BH formation. We take inspiration from works that analyse the prompt emission as a phase when BH forms and grows and extrapolate the results to the time when the magnetically arrested disk (MAD) forms and the delivery of the angular momentum to BH via accretion becomes inefficient. We assume further glow via the \cite{BlandfordZnajek} process and solve numerically the energy and angular momentum conservation equations. Those assumptions lead to a simple model of a BH spin-down, powering the afterglow emission. 
Although in the literature there has been discussion in various contexts of the BH spin-down for GRBs \citep{vanPutten2004,KumarZhang2015,Lei_2017,Li2018b,vanPutten2023,Issa2025,Wu2025}, it is the first time, to the best of our knowledge, that this mechanism is discussed as a primary physical process driving the plateau and showing the $L_{X}$ - $T^*_{X}$ anti-correlation. Our results predict the existence of a $L_X\sim T^{*\,-1}_{X}$ relation with scatter driven by the varying mass and the initial spin of the BHs. We discuss those results in the context of observed differences between SGRBs and LGRBs, as well as the redshift evolution of GRBs.
Some of us found in \cite{Dainotti2020ApJ...904...97D} that all kilonova-associated GRBs have statistically dimmer plateaus than LGRBs. The previous analysis shows that all those GRBs fall below the $L_{X}$-$T^*_{X}$-$L_{peak}$ plane obtained for LGRBs. This behaviour remained a mystery for almost 5 years. Remarkably, the work outlined in this manuscript finally allowed us to provide a theoretical interpretation as to why such a phenomenon occurs.
In Sec. \ref{sec:datasampe} we discuss how the data was collected from different instruments. In Sec. \ref{sec:analysis} we explain how every considered parameter is estimated, perform the analysis of the cosmological evolution of parameters and analyse the correlations. In Sec. \ref{sec:BH} we present the simple BH spin-down model. Sec. \ref{sec: shock re-emission} discusses the possible physical explanation for the 3D $L_{X}$ - $T^*_{X}$ - $L_{opt}$ correlation. Finally, in Sec. \ref{sec:conclusions} we summarize our work and draw conclusions. 

\section{Data sample}
\label{sec:datasampe}

\subsection{X-ray afterglows}

Our sample of GRB afterglows consists of sources with known redshift observed by {\it Swift} since its launch in 2005 until February 2024 (427 sources). We analysed all events observed by both the Burst Alert Telescope (BAT) and X-ray Telescope (XRT) instruments, selecting 255 GRBs with known redshift exhibiting a clear plateau phase presence in their LCs.
The BAT is designed to detect GRBs in the 15-150 keV energy range. It can rapidly identify GRBs, triggering {\it Swift}'s follow-up observations. Once a burst is detected by BAT, the XRT is used to study the GRB's afterglow in the 0.3-10 keV energy range, providing detailed X-ray LCs and spectra. The XRT is capable of both rapid slewing to the GRB's location and continuous monitoring of its afterglow for extended periods, making it compelling for tracking the evolution of the GRB afterglow, including the plateau phase.

\subsection{High-energy peak emission}

\cite{Dainotti2016} discovered a tight log-linear correlation between the X-ray afterglow properties (rest-frame time of the end of the plateau $log_{10}(T^*_X)$ and the corresponding luminosity $log_{10}(L_X)$), and then extended the correlation to 3D by involving the high energy prompt luminosity obtained over a $\sim 1\rm s$ interval after the trigger ($log_{10}(L_{\rm peak})$). The addition of a third dimension allowed to significantly reduce the scatter of the relation. Further proved by \cite{CaoShulei2022}. Thus, our study needs to gather not only the best quality afterglow but also prompt data. We focus further on the $L_{\rm peak}$ observed by either {\it Swift}'s BAT ($L_{\text{peak, {\it Swift}}}$) or {\it Fermi}'s Gamma-ray Burst Monitor (GBM) ($L_{\text{peak, Fermi}}$). 
GBM observes the prompt emission of GRBs in the 10-10000 keV energy range, thus highly extending the spectral coverage of {\it Swift}. It provides key measurements such as the burst duration, $T_{90}$, the peak flux $F_{\text{peak, Fermi}}$ (measured over 1024 ms), the fluence $f_{\text{Fermi}}$, and the best-fit spectral parameters. The burst spectra are typically fitted using the three-parameter \cite{Band1993} function, characterized by the break energy ($E_{\text{p}}$), the low-energy slope ($\alpha$), and the high-energy slope ($\beta$). 
This empirically introduced function can be constructed within the model of photospheric emission \citep{Thompson1994,Eichler2000,Mészáros2000,Rees2005,Lazzati2009,Pe'er_2011,Mizuta2011,Nagakura2011,Ruffini_2013,Xu_2012,Bégué_2013,Lundman2013,Lazzati_2013}, from the structured jet \citep{Ito2014}. Alternative explanation for the shape of the spectra is the efficient dissipation around the photosphere \citep{Pe’er_2005,Pe’er_2006,Giannios2006,Giannios2008,Giannios2007,Ioka2007,Lazzati2010,Beloborodov2011,Vurm2011,Asano2013}. The photospheric emission model can explain the \cite{Yonetoku2004} relation, see \citep{Ito2014} for details.
In some cases, the parameters $\beta$ and/or $E_{\text{peak}}$ cannot be well constrained, simplifying the spectral model to either a power law with a high-energy cut-off or a simple power law. We have compiled a collection of best-fit spectral parameters for 201 GRBs with known redshifts, all observed by {\it Fermi} to date, providing a thorough investigation for analyzing the prompt emission characteristics of these bursts.

\subsection{Optical afterglows}

Optical observations of GRBs pose more challenges than X-ray observations, primarily because the ground-based telescopes must be pointed accurately at the burst's location within seconds of the trigger. A high fraction of bursts are detected by the Ultraviolet Optical Telescope (UVOT) aboard the {\it Swift} satellite. However, this instrument's limited sensitivity constrains the number of successful observations. As a result, many GRBs remain undetected in the optical band by {\it Swift} alone.
To address this limitation, \cite{Dainotti2022opt} compiled data from various other facilities to create a comprehensive sample of optical observations. By combining detections from different telescopes, they provided an enriched sample. For all LCs, the measurements taken through different filters were rescaled to the same band. The authors provided the community with 179 sources in R-band flux that exhibit the plateau phase with known redshift and the spectral power-law index taken from the literature. \cite{Dainotti2022opt} assumed that there is no spectral evolution during the afterglow phase. This is a pretty reliable approximation since $\sim 85.5\%$ of the cases investigated in \cite{Dainotti2024opt} with sufficient data coverage show no colour evolution. Assessing the non-variability of the SED is crucial to determine if the rescaling of the different filters can be performed.
In some instances, the situation is further complicated by insufficient optical data to precisely determine the SED shape. When this occurs, the authors extrapolate the necessary information from the X-ray data to complete the calculations. 
A more detailed discussion of the 
optical data used in this study can be found in \citet{Dainotti2022opt}.

\subsection{Joined sample}

\begin{figure}[ht!]
    \centering
    \includegraphics[width=0.99\linewidth]{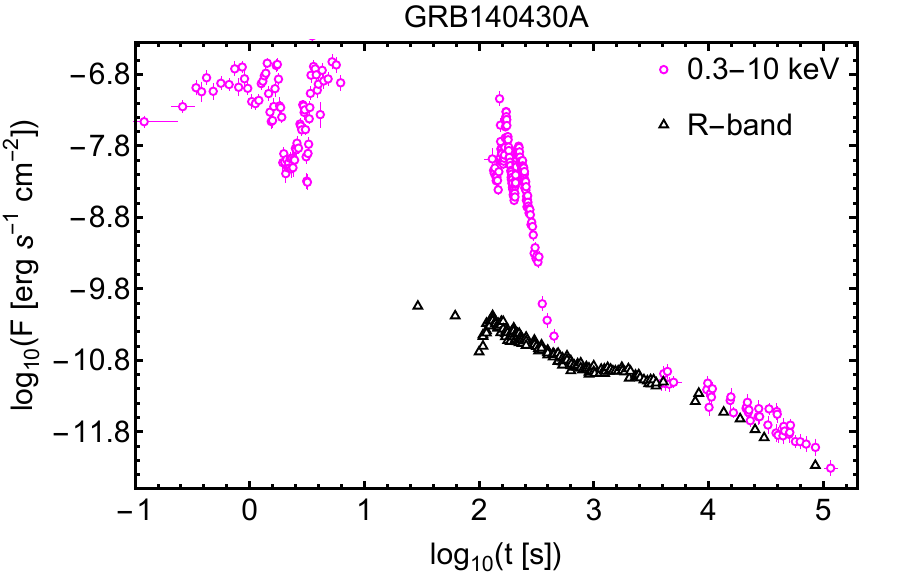}
    \caption{The figure presents overplotted X-ray (magenta circles) and optical (black triangles) data for the light curve of GRB 140430A. We present the observed flux data vs observer-frame time. This is raw data before any rest-frame transformation.}
    \label{fig:140430aLC}
\end{figure}

We cross-matched the sources observed by both {\it Swift} and {\it Fermi} with the {\it Swift} data showing an X-ray plateau. We obtained a set of 78 sources with a redshift range of 0.0093 - 8. The {\it Swift} XRT and optical ({\it Swift}-UVOT + ground-based) catalogues have an overlap of 52 sources spanning from $z=0.15$ to $z=5.9$, while {\it Fermi} GBM and optical have an overlap of 48 sources with a redshift range of 0.15-8. For clarity, we present separate sample sizes in Table \ref{tab:overlap}. We demonstrate an example of the over-plotted X-ray and optical LCs for GRB 140430A in Fig. \ref{fig:140430aLC}.

\begin{table}[ht!]
    \centering
    \begin{tabular}{|c||c|c|c|}
    \hline
        N & {\it Swift} XRT & {\it Fermi} GBM & Optical All \\\hline\hline
        {\it Swift} XRT & 255 & 78 & 52 \\\hline
       {\it Fermi} GBM & 78 & 201 & 48 \\\hline
        Optical All & 52 & 48 & 179 \\\hline
    \end{tabular}
    \caption{Table presenting the number of GRBs with overlapping observations from three different catalogues ({\it Swift} XRT, {\it Fermi} GBM, and optical).}
    \label{tab:overlap}
\end{table}

\section{Analysis}
\label{sec:analysis}

To ensure the statistical significance of the plateau, we used selection criteria based on the Akaike Information Criterion (AIC). For each afterglow, we determine the time of the end of the plateau $T_{X}$ and the corresponding flux $F_{X}$ (observed in X-rays) as the best-fit parameters of the W07 function. We performed least squares fitting using two models: a simple power law and the \cite{Willingale2007} (W07) model:

\begin{equation}
    F(t) = \left\{ \begin{matrix}
        F_{X} \times e^{\alpha(1-\frac{t}{T_X})}, & t<T_X \\
        F_{X} \times (\frac{t}{T_X})^{-\alpha}, & t\geq T_X .
    \end{matrix}\right. 
\end{equation}

AIC values were calculated for both models and the likelihood ($\mathcal{L}_{W07}$) that the W07 model is favoured in relation to a simple power-law (PL) was determined using the following sequence of equations:

\begin{equation}
    B_{W07}=exp((min(AIC)-AIC_{W07})/2),
\end{equation}
\begin{equation}
    B_{PL}=exp((min(AIC)-AIC_{PL})/2),
\end{equation}
\begin{equation}
    \mathcal{L}_{W07}=\frac{B_{W07}}{B_{W07}+B_{PL}}.
\end{equation}

We define a well-visible plateau as one for which $\mathcal{L}_{W07}>0.95$. This sample builds on a previous work of some of us \citep{Srinivasaragavan2020}, expanding the analysis with a larger set of GRBs and a more rigorous method for plateau detection. This analysis depends on the choice of model for the fitting. However, we obtained a very similar sample when fitting the afterglows with the broken power law or the "ideal magnetar" ($F(t)=F_0\times (1+t/T)^{-2}$ \cite{Wang2022ApJ...924...97W, Wang2024}) models. Thus, our treatment provides a reliable sample of afterglows characterized by the presence of the plateau. We chose the sample for the W07 formula since it is the largest set obtained. Starting with 427 GRBs with known redshift observed by {\it Swift} from its launch in 2005 up to February 2024,  60\% (255 GRBs) survived these criteria.

Further, we employ the observed redshift to compute the rest frame time at the end of the plateau $T^{*}_{X}$. Under the assumption of a flat $\Lambda$CDM cosmological model ($\Omega_M=0.3$ and $H_{0}=70\,\rm\frac{km}{s\, Mpc}$), we calculate the luminosity corresponding to this moment:

\begin{equation}
    L_X=F_X\times 4 \pi d^{2}_{L}(z)\times K(\beta,z),
\end{equation}

where $d_{L}(z)$ is the luminosity distance computed for the redshift $z$, $K(\beta,z)$ is the so-called K correction \cite{Bloom2001}, providing us with a transformation of luminosities to the source rest-frame band. This parameter is computed under the assumption of power-law spectra with photon index $\beta$, determined from the {\it Swift} observations. The same treatment is performed for the optical data and results in obtaining the rest-frame time of the end of the optical plateau $T^*_{opt}$ and the corresponding optical luminosity $L_{opt}$. Both $L_{peak,\, Swift}$ and $L_{peak,\, Fermi}$ are obtained from energy fluxes observed by corresponding telescopes throughout $\sim 1 s$ from the burst trigger in the instrument's frame. Moreover, we define $E_{X}$ and $E_{opt}$ parameters as the energy emitted during the plateau in X-ray and optical bands, respectively. 
We estimate their values as: $E_{X} = L_{X}\times T^{*}_{X}$, $E_{opt} = L_{opt}\times T^{*}_{opt}$.

\subsection{Correction for evolution}
\label{sec:evolution}

It is well-established that GRBs exhibit cosmological evolution, meaning their properties change as a function of redshift. For instance, GRBs from the early universe are generally observed to be brighter than those from later epochs. Such a redshift dependence can artificially drive correlations between various GRB parameters. To address this, we aim to remove the redshift dependence by modelling it with an assumed functional form. Specifically, we assume that the redshift-dependent distribution of a given parameter $X(z)$ can be expressed as the product of its distribution at $z=0$ (X') and an evolutionary function $\delta(z) = (1+z)^\kappa$:

\begin{equation}
    X(z) = X' \times \delta(z).
\end{equation}

This is the simplest approach to modelling evolution. However, determining the coefficient $\kappa$ is non-trivial due to selection biases that can introduce artificial correlation. Therefore, it is essential to use robust statistical methods to quantify the relationship. A commonly used method in the literature is the statistical approach of \cite{EffronPetrosian1992}, which examines correlations within subsets of the parameter space truncated only along the x and y axes, avoiding artificial correlations. For more details, see \ref{Appendix}.
In this work, we apply this method to the $L_{\text{peak, Fermi}}$, $E_{\rm iso}$ and $E_{\rm peak}$ parameters. The evolution of $L_{peak,\, Swift}$, $L_X$, $T^*_{X}$, was first obtained by 
\cite{Dainotti2013,Dainotti2015,Dainotti2017a,dainotti2017} and more recently by \cite{Dainotti2020ApJ...904...97D}, while for the $L_{opt}$, and $T*_{opt}$ parameters by  \cite{Dainotti2022opt}. The evolutionary coefficients used in this work: 
\begin{itemize}
    \item $\kappa_{L_{X}} = 2.42\pm 0.58 $ - evolutionary parameter determined for $L_X$ for the sample of 222 GRBs
    \item $\kappa_{T^{*}_{X}} = -1.25 \pm 0.28 $ - evolutionary parameter determined for $T^{*}_X$ for the sample of 222 GRBs
    \item $\kappa_{L_{\text{peak, Swift}}} = 2.24\pm 0.30 $ - evolutionary parameter determined for $L_{\text{peak, Swift}}$ for the sample of 222 GRBs
    \item $\kappa_{L_{opt}} = 3.96 \pm 0.43 $ - evolutionary parameter determined for $L_{opt}$ for the sample of 180 GRBs
    \item $\kappa_{T^{*}_{opt}} = -2.11\pm 0.49 $ - evolutionary parameter determined for $T^{*}_{opt}$ for the sample of 180 GRBs
    \item $\kappa_{L_{\text{peak, Fermi}}} = 2.76\pm 0.34 $ - evolutionary parameter determined for $L_{\text{peak, Fermi}}$ for the sample of 201 GRBs
    \item $\kappa_{E_{\text{iso, Fermi}}} = 2.18 \pm 0.55 $ - evolutionary parameter determined for $E_{\text{iso,Fermi}}$ for the sample of 201 GRBs
    \item $\kappa_{E_{\text{peak, Fermi}}} = 0.72 \pm 0.26 $ - evolutionary parameter determined for $E_{\text{peak, Fermi}}$ for the sample of 135 GRBs (the number is smaller than the full sample because we determine $E_{\rm peak}$ only if $\beta<-2$ (the high energy slope of spectra).
    
\end{itemize}

\subsection{Fermi vs {\it Swift}}
\label{Fermi-Swift}

\cite{dainotti2017} showed with a sample of 34 LGRBs that the scatter of the fundamental plane is reduced by employing the {\it Fermi} GBM prompt data instead of {\it Swift}'s BAT. However, at that time, with that small sample, it was challenging to draw definitive conclusions. In this work, we have more than doubled the sample for the prompt data from the GBM telescope and we have gathered 78 GRBs observed by both {\it Swift} and {\it Fermi}, which were characterized by the presence of the X-ray plateau. To ensure correspondence between the two observations, we cross-matched the data both in terms of the time of the trigger and position in the sky. Furthermore, we segregated GRBs into long and short based on the {\it Fermi} $T^*_{90}=2s$ boundaries and controlled the data quality by dropping the cases where the uncertainty on the peak luminosity was greater than the luminosity itself. This limit on $T^*_{90}$ is likely an overestimate, as we will show further in the article, but to ensure that all considered GRBs belong to the long class, we choose it as a primary selection. The output provides us with a set of 61 high-quality LGRBs with a high probability that two different satellites observed the same event. 
We compare the parameters of the Dainotti relation fitted to different data sets in Tab. \ref{tab:Fermi}. We consider both cases with and without correction for evolution and reference previous results from \cite{dainotti2017}. One can notice that the same set of GRBs fitted without correction for evolution with {\it Fermi} prompt data has $\sim 30\%$ lower scatter ($\sigma_{\text{int, Fermi}}=0.32\pm0.03$) than in the case of the use of {\it Swift} prompt data ($\sigma_{\text{int, Swift}}=0.46\pm0.05$). The z-score computed between the two is $\sim 2.4$, a significant improvement in results by \cite{dainotti2017} (reduction by z-score: $\sim 0.5$). Including correction for evolution allowed us to reduce scatter in both cases further. For the {\it Fermi} data, we obtained $\sigma_{\text{int, Fermi, EV}}=0.25\pm0.04$, which, compared to the {\it Swift} data with correction for evolution ($\sigma_{\text{int, Swift, EV}}=0.41\pm0.05$), is a $\sim 39\%$ reduction, and the z-score between the two estimates is $\sim 2.5$.


\begin{table*}[ht!]
    \centering
    \begin{tabular}{|c||c|c|c|c|c|c|}
    \hline
        Sample & a & b & c & $\sigma_{int}$ & N & $R^2$ \\\hline\hline
        Long GRBs (Fermi) & $-0.85\pm0.09$ & $0.68\pm 0.05$ & $13.6\pm 3.0$ & $0.32\pm 0.03$ & 61 & 0.90 \\\hline
        Long GRBs with correction for evolution (Fermi) & $-0.76\pm0.10$ & $0.73\pm 0.07$ & $10.3\pm 3.7$ & $0.25\pm 0.04$ & 61  & 0.89 \\\hline
        Long GRBs ({\it Swift}) & $-0.89\pm0.12$ & $0.80\pm 0.10$ & $9.6\pm 5.2$ & $0.46\pm 0.05$ & 61  & 0.82 \\\hline
        Long GRBs with correction for evolution ({\it Swift}) & $-0.84\pm0.13$ & $0.80\pm 0.13$ & $9.5\pm 6.6$ & $0.41\pm 0.05$ & 61  & 0.81 \\\hline
        Long GRBs (Fermi) \citep{dainotti2017} & $-0.89\pm0.07$ & $0.58\pm 0.10$ & $21.34\pm 5.96$ & $0.43\pm 0.07$ & 34  & - \\\hline
        Long GRBs ({\it Swift}) \citep{dainotti2017} & $-0.88\pm0.09$ & $0.65\pm 0.13$ & $17.22\pm 7.50$ & $0.48\pm 0.07$ & 34  & - \\\hline
    \end{tabular}
    \caption{Table presents results of the fitting of the $log_{10}(L_{X}) = a\times log_{10}(T^{*}_{X}) + b\times log_{10}(L_{\rm peak}) + c$ correlation to the sample of LGRBs in two cases: the peak luminosity estimated from {\it Fermi} data $L_{\rm peak}=L_{\text{peak, Fermi}}$, and the peak luminosity estimated from {\it Swift} data $L_{\rm peak}=L_{\text{peak, Swift}}$. The $\sigma_{int}$ denotes the intrinsic scatter estimated in each case, and N is the number of sources in the studied sample. We present results with and without correction for evolution. The last two rows present the estimates obtained by \protect \cite{dainotti2017}. }
    \label{tab:Fermi}
\end{table*}

\subsection{Collapsar or Merger?}
\label{sec:clustering}

\subsubsection{The Dainotti relation and the $E^*_{\rm peak}$ - $E_{\rm iso}$ Diagram}

\begin{figure}[ht!]
    \centering
    \includegraphics[width=0.99\linewidth]{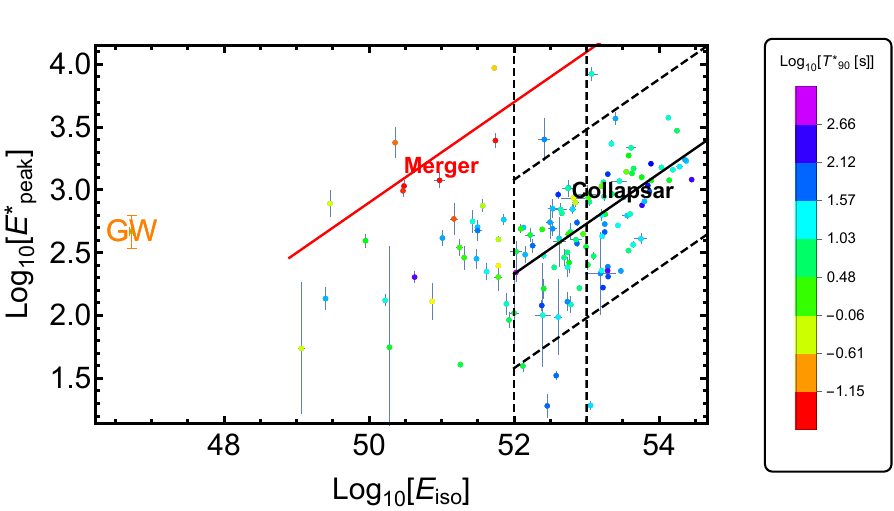}
    \caption{The $log_{10}(E^*_{\rm peak}\, \rm [keV])$ - $log_{10}(E_{\rm iso}\, \rm [erg])$ diagram constructed for 201 {\it Fermi} GRBs with known redshift. The black, continuous line marks the Amati correlation with slope $\sim 0.4$, and the dashed black lines perpendicular to the continuous one mark the $0.45\,\rm dex$ intervals from the Amati correlation. The vertical dashed black lines mark the energy limits $10^{52}$ erg - total kinetic energy of a standard $1.4\, M_{\odot}$ magnetar, and $10^{53}$ - total kinetic energy of theoretical the most extreme possible magnetar.}
    \label{fig:DivEpeakEiso}
\end{figure}

\cite{Zhang2009} performed a detailed classification of GRBs into the merger and collapsar origin classes, based on their duration, host galaxy, circumburst environment, and Supernova association. The authors investigated the properties of the samples and constructed the $E^*_{\rm peak}$ - $E_{\rm iso}$ diagram, where $E^*_{\rm peak}=E_p\times(2+\alpha)\times(1+z)$ (for $\beta<-2$) is the rest-frame peak energy measured from the time-integrated $\nu F_{\nu}$ spectra, and $E_{\rm iso}$ is the isotropic-equivalent energy emitted during the prompt phase \citep{Schaefer2007}. The authors noticed the clustering of the two sets in different regions of the parameter space. The finding was confirmed through multiple studies in the literature. Recently \cite{Minaev2020}, gave another recipe on how to classify GRBs using the $E^*_{\rm peak}$ - $E_{\rm iso}$ diagram and $T^*_{90}$ only.
Following this analysis, we start our analysis by constructing the $E^*_{\rm peak}$ - $E_{\rm iso}$ diagram with $T^*_{90}$ marked with colour for all GRBs observed by {\it Fermi} (see Fig. \ref{fig:DivEpeakEiso}).

One can pinpoint two well-known clusters of data from the literature, studied in the aforementioned papers. The group of LGRBs (bursts of collapsar origin) following the Amati correlation (shown as a continuous black line) \cite{amati2006} form a highly populated region with $E_{\rm iso}>10^{52} erg$ and $T^*_{90}>2\,\rm s$. We mark the approximate borders of this region with dashed black lines.
Above the LGRB cluster, we see SGRBs (bursts of NS-NS merger origin) which also have smaller values of $E_{\rm iso}$. Additionally, below $E_{\rm iso}=10^{52} erg$, there is present a more dispersed cluster of GRBs with $T^*_{90}>2s$. Some of those still follow an extension of the LGRBs or SGRBs regions. The clustering present in this diagram is commonly used in the literature as a powerful tool for discriminating the origin of GRBs. \cite{Minaev2020} defined a new parameter $EH$ - energy-hardness, which is a rescaled residual of the Amati correlation with the slope $\sim 0.4$. We define this parameter as:

\begin{equation}
    EH = log_{10}(E^*_{\rm peak}\text{ [keV]}) - 0.40 \times log_{10}(E_{\rm iso}\text{ [erg]}) + 18.47 .
    \label{eq:EH}
\end{equation}

Further, the authors demonstrated how such parametrization can be applied to pinpoint the class of a given burst and how robust the method is.
A similar idea about the residual appeared in the literature much earlier, in  \cite{DelVecchio2016}, where they observed that the residuals of the $L_X$ - $T^*_X$ relation show a correlation with the slope of the plateau. 
We take inspiration from this analysis and study the distribution of sources across the 3D Dainotti relation. By analogy to \cite{DelVecchio2016} and a similar nomenclature of \cite{Minaev2020}, we define the parameter "Plateau Shift" (PS), i.e., the residual of the 3D Dainotti relation, as:

\begin{equation}
\resizebox{0.45\textwidth}{!}{%
    $PS = log_{10}(L_X)+0.85\times log_{10}(T^*_X)-0.68\times log_{10}(L_{Fermi,\, peak})-13.6.$%
    }
    \label{eq:PS}
\end{equation}

\begin{figure*}[ht!]
    \centering
        \includegraphics[width=0.49\linewidth]{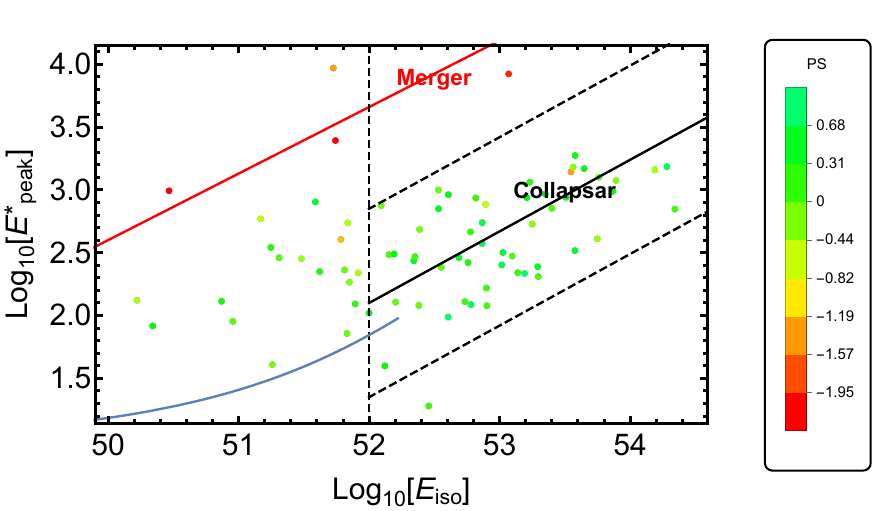}
    \includegraphics[width=0.49\linewidth]{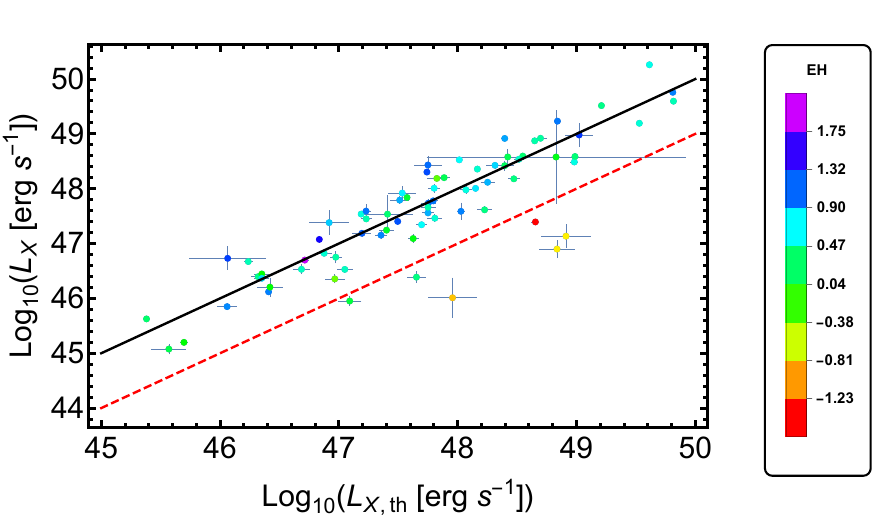}

    \caption{Left: The $log_{10}(E^*_{\rm peak}\, \rm [keV])$ - $log_{10}(E_{\rm iso}\, \rm [erg])$ diagram constructed for 78  GRBs with known redshift observed by both {\it Swift} and {\it Fermi}. The black, continuous line marks the Amati correlation with slope $\sim 0.4$, and the dashed black lines perpendicular to the continuous one mark the $0.75\,\rm dex$ intervals from the Amati correlation. The blue line marks the border on the non-observable region by considered satellites due to Malmquist bias. The colour marks the residuals from the FP correlation. Right: the plot of observed vs predicted by the Dainotti relation $log_{10}(L_X)$ values. The colour marks the EH parameter. The dashed red line marks the 1 dex offset from the equality line.}
    \label{fig:ResEisoEpeak}
\end{figure*}

Where the numerical coefficients were obtained as best-fit plane parameters for the 61 LGRBs (First row of Tab.\ref{tab:Fermi}). In the right panel of Fig. \ref{fig:ResEisoEpeak}, we present the corresponding plot to Fig. \ref{fig:DivEpeakEiso}, but constructed only for GRBs observed contemporaneously by both {\it Swift} and {\it Fermi} and with PS marked as colour. 

One can easily notice that almost all sources with highly overestimated values of $log_{10}L_{X,\, th}$ (laying under the plane of correlation) fall in the region previously identified as populated by merger-type events. The reverse situation is presented in the left panel of Fig. \ref{fig:ResEisoEpeak}. We present the observed $L_X$ as a function of the predicted one by the Dainotti correlation. We observe a prominent cluster of sources with a small EH parameter that falls significantly under the plane of correlation.

\subsubsection{EH - PS diagram}

\begin{figure}[ht!]
    \centering
    \includegraphics[width=0.99\linewidth]{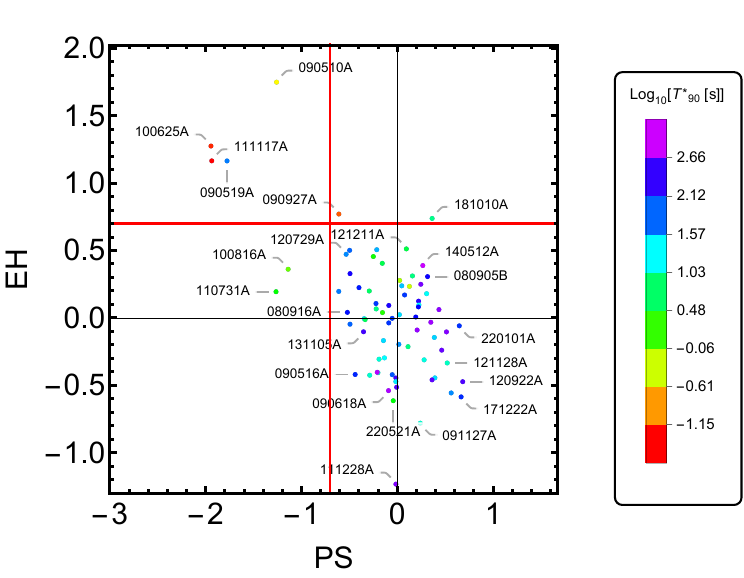}
    \caption{The diagram of EH as a function of PS. The color marks the values of $log_{10}(T^*_{90})$. The horizontal red line marks $EH=0.7$, while the vertical one $PS=-0.7$. We interpret those lines as borders between LGRBs and SGRBs.}
    \label{fig:EHPS}
\end{figure}

To determine the borders between different classes of data, we construct the Energy Hardness - Plateau Shift diagram. The plot is illustrated in Fig. \ref{fig:EHPS}.

One can notice a well-defined cluster of points with $-0.7>PS>0.7$ and $-0.7>EH>0.7$. All sources present in this region have $T^*_{90}>1\, \rm s$. Outside of this cluster, one can observe the presence of bursts with $T^*_{90}<2\,\rm s$, which do not cluster in an organized way. 

\subsubsection{Are all the outliers of FP Short GRBs?}

Here we point all GRBs laying at least 0.7 dex below the plane of Dainotti relation, and discuss what is their origin in reference to the literature:
\begin{itemize}
    \item GRB090510A $T^*_{90} = 0.51\pm0.07\rm s$, $z=0.903$. 
    \cite{Yuan2021} discussed a potential merger origin of this GRB, which would result in a KNe.
    On the other hand, \cite{Muccino2013} proposed that this GRB is likely to be an LGRB disguised as an SGRB; however, in a model-dependent manner, which requires a rather extreme value of the circumburst medium density ($\sim 2\times 10^3\rm cm^{-3}$) required to fit the data. The authors concluded that even if this is, in fact, LGRB, it belongs to a new class of GRBs originating in very dense galaxies.
    \item GRB 090519A $T^*_{90} = 15.2\pm1.1\rm s$, $z=3.85$. 
    This GRB was never the subject of a single-source study. However, we observe an extreme value of the spectral peak energy $E^*_{\rm peak}\approx 10^4\rm keV$, usually observed in SGRBs. Moreover, the peak energy is very close to the observable limit of the BAT spectrograph, which could distort the estimated parameters and lead to misclassification. We conclude that this is most likely a merger-origin GRB.
    \item GRB100625A $T^*_{90} = 0.17\pm0.19\rm s$, $z=0.452$.
    This GRB was never questioned for its affiliation with the SGRB class. \cite{Yuan2021} discussed a potential merger origin of this GRB, which would result in a Kilonove. According to the authors, such an emission could be observed by the most sensitive contemporary instruments like the Vera C. Rubin Observatory.
    \item GRB100816A $T^*_{90} = 1.13\pm0.13\rm s$, $z=0.804$.
    \cite{Tunnicliffe2014} discusses that this event is an SGRB due to its position within the host, spectral hardness, and the lack of associated SNe.
    \item GRB110731A $T^*_{90} = 1.95\pm0.15\rm s$, $z=2.83$.
    This is a widely studied source in the literature since it is the first GRB observed in GeV, X-ray, and optical bands with a good data coverage \citep{Ackermann2013}. \cite{Lu2017} was the first to analyse a possible progenitor of this event. The authors concluded that the magnetar scenario for this GRB requires extreme properties of an NS (although still physically possible; magnetic field: $B\leq 10^{16}\,\rm G$ and the initial period of newborn magnetar: $P_0\leq0.56\,\rm ms$), see \citep{dallosso2018}. Moreover, this source has a large physical distance from its host galaxy. Thus, this event is most likely a result of a compact binary merger. An alternative explanation is provided by \cite{Primorac2018}, where this event is classified as an LGRB within the Binary-driven Hypernova scenario. However, this result is unlikely due to the misalignment of GRB and the host galaxy. As with all SGRBs, this source falls significantly under the plane of the Dainotti relation, and there is no physical explanation of this fact given by the Binary-driven Hypernova scenario. 
    \item GRB 111117A $T^*_{90} = 0.135\pm0.026\rm s$, $z=2.211$.
    Due to the high precision of the duration estimate, this GRB is always classified in the literature as an SGRB \citep{2018A&A...620A..37C, 2018A&A...616A..48S, 2013ApJ...766...41S, 2012ApJ...756...64Z, 2012ApJ...756...63M, 2012grb..confE..73S}.
\end{itemize}

Summarizing, we find good evidence that all GRBs falling significantly under the plane of $log_{10}L_X$ - $log_{10}T^*_X$ - $log_{10}L_{\rm peak}$ correlation (PS < -1) belong to the merger origin class. The results presented above do not depend on the choice of the instrument which measures the $L_{\rm peak}$. We also stress here that the same results are obtained with correction for evolution applied to all studied parameters.

\subsubsection{$E_{\rm iso}$ - $T^*_{90}$ diagram}

\begin{figure}[ht!]
    \centering
    \includegraphics[width=0.99\linewidth]{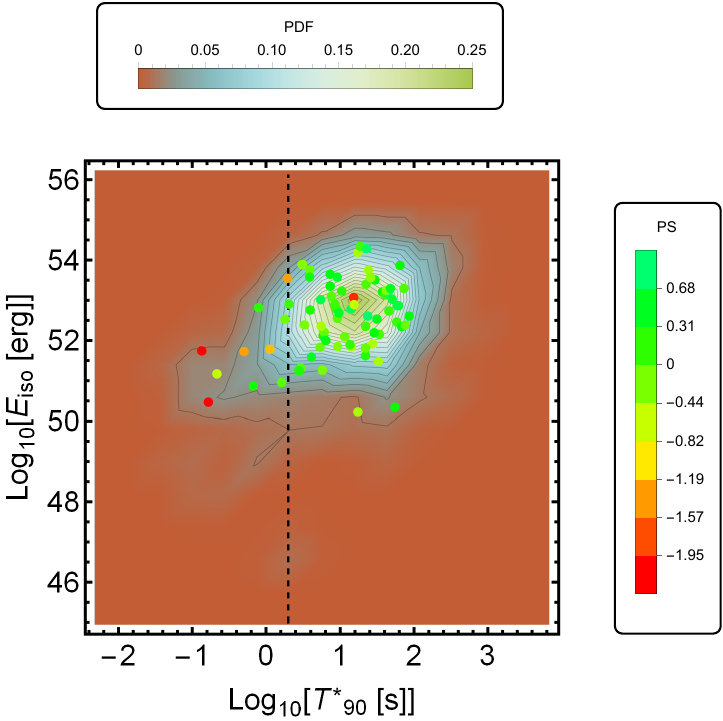}
    \caption{The distribution of a sample of GRBs observed by both {\it Swift} and {\it Fermi} (points) over-plotted on the distribution (PDF) of all GRBs observed by {\it Fermi} (the shades of orange, blue and green) in the $log_{10}(E_{\rm iso})$ - $T^*_{90}$ space. The colours of the points mark the value of the PS parameter. The dashed black line corresponds to $T^*_{90}=2\,\rm s$.}
    \label{fig:ResEisoT90}
\end{figure}

Furthermore, we mark the PS as a colour in the $E_{\rm iso}$ - $T^*_{90}$ diagram. We present such a plot in Fig. \ref{fig:ResEisoT90}.

As one can notice, the traditional division of GRBs into long and short is not strong since the SGRBs (red and orange points) are dispersed in the considered space. On the density plot, we see a clear arm that extends from a distribution of LGRBs towards small values of $T^*_{90}$, which is composed mainly of the S-GRBs.

\subsection{Dependence of the results on redshift}

For the reliability of results, we carefully investigated selection biases and redshift evolution to see if this can induce the clustering. We computed the non-observable region of the $log_{10}(E^*_{peak})$ - $log_{10}(E_{iso})$ diagram as a parametric function obtained for the minimum observable fluence and $E_{peak}$ of the {\it Fermi} GBM instrument for the redshift range 0-9. The obtained line is marked with blue colour in Fig. \ref{fig:ResEisoEpeak}. As visible, the selection bias has negligible influence on the distribution of sources at $E_{iso}>10^{52}\,\rm  erg$. We stressed in Sec. \ref{sec:evolution} that properties of GRBs change with redshift, thus it is also crucial to test if the clustering might be due to evolution. We repeated the whole above analysis for the de-evolved parameters: $E'_{iso}=E_{iso}/(1+z)^{\kappa_{E_{iso}}}$, $E^{*'}_{peak}=E^*_{peak}/(1+z)^{\kappa_{E^*_{peak}}}$, $L'_{X}=L_{X}/(1+z)^{\kappa_{L_{X}}}$, etc., and we obtained the same results in terms of belonging to collapsar and merger class. Namely, the outliers of the Amati relation identified as SGRBs are still outliers, and the same holds for the Dainotti relation. Therefore, we conclude that our analysis of clustering of SGRBs and LGRBs is independent of both selection bias and redshift evolution.

\subsection{Searching for additional plateau correlations}
\label{sec:LxLo}

\begin{figure*}[ht!]
    \centering
    \includegraphics[width=0.75\linewidth]{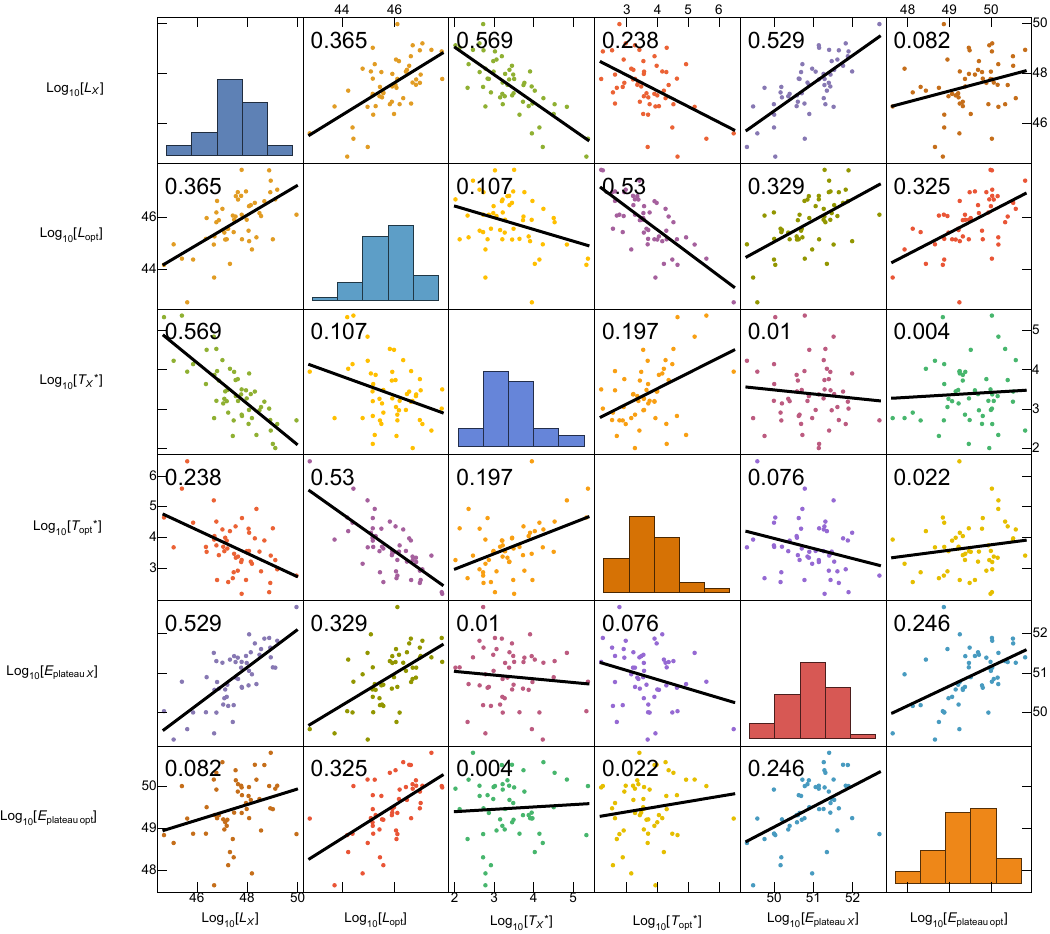}
    \caption{A scatter matrix plot, presenting the correlations between different parameters observed in X-ray and optical. The over-plotted numerical values indicate Pearson's $R^2$ coefficient.}
    \label{fig:CorSearch}
\end{figure*}

We present the empirical relations between X-ray ($L_{X}$, $T^*_X$, $E_X$), and optical ($L_{opt}$, $T^*_{opt}$, $E_{opt}$) parameters in Fig. \ref{fig:CorSearch}. This compilation of plots suggests, that despite known correlations ($L_{X}-T^{*}_{X}$ and $L_{opt}-T^{*}_{opt}$ \citep{Dainotti2008,Dainotti2020}), there seems to exist a residual nonlinear correlation between $L_{X}$ and $L_{opt}$. Moreover, a weak non-linear correlation exists between the energy emitted during the X-ray plateau and the corresponding optical energy.

\begin{figure*}[ht!]
    \centering
    \includegraphics[width=0.75\linewidth]{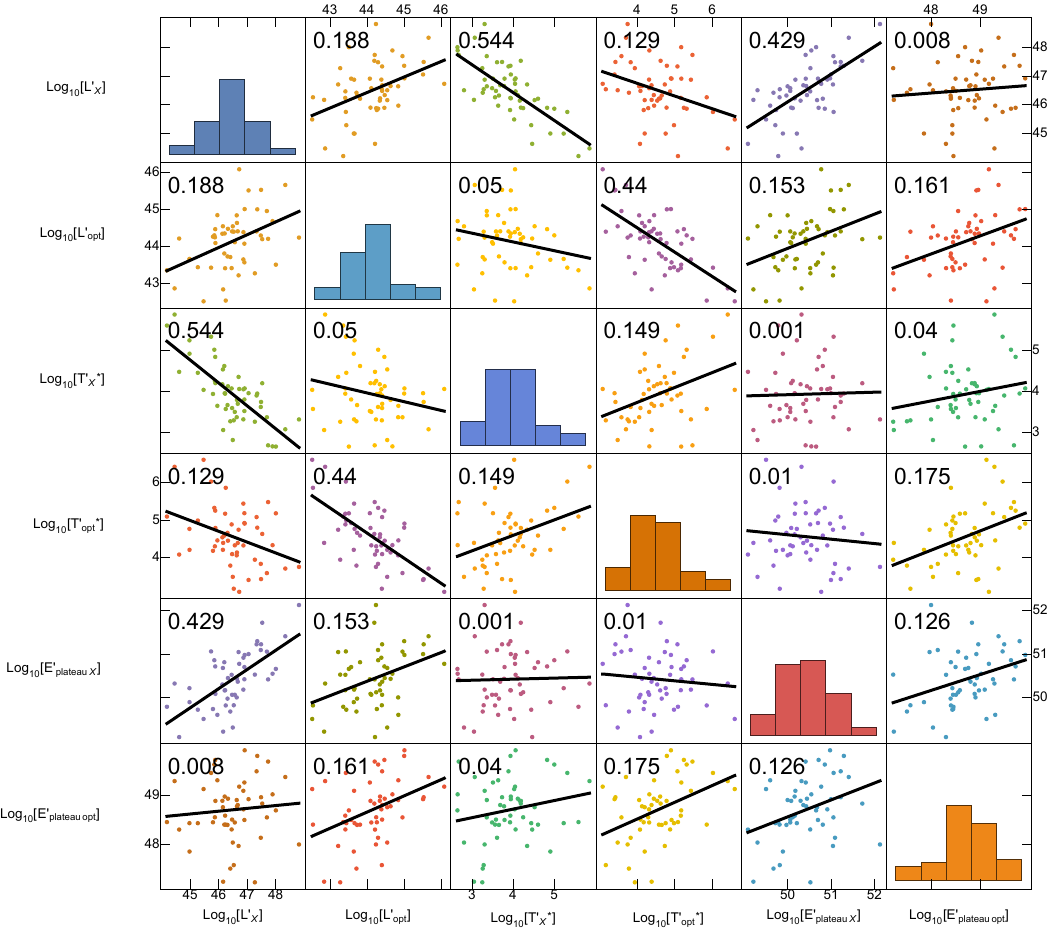}
    \caption{A scatter matrix plot, presenting the correlations between different parameters observed in X-ray and optical, and corrected for evolution. The over-plotted numerical values indicate Pearson's $R^2$ coefficient.}
    \label{fig:CorSearchEv}
\end{figure*}

As we discussed in Sec. \ref{sec:evolution}, the evolution of data with redshift can artificially induce a correlation. Some examples of artificially induced correlation can be found in the literature (see \cite{DainottiAmati2018} and \cite{Dainotti2017a} for an overview). Therefore, it is crucial to test the de-evolved data. We present the scatter plot of de-evolved parameters in Fig. \ref{fig:CorSearchEv}.

\begin{figure}[ht!]
    \centering
    \includegraphics[width=0.99\linewidth]{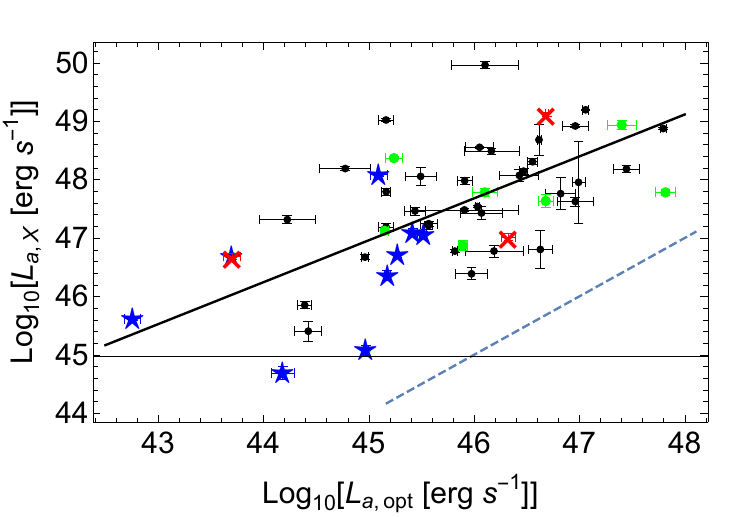}
    \caption{Plot of luminosity at the end of the plateau measured in X-rays as a function of the corresponding luminosity measured in optical R band. The style of plot markers distinguishes the class of a given GRB. Long GRBs are presented as black circles, short as red crosses, SNe associated as blue stars, and XRRs, together with XRFs, are marked as green circles. The black solid line is a simple linear regression best fit. The region below the dashed blue line is the unobservable region computed with the assumption of minimum observable fluxes: $F_{X,\, min} = -12.5\,\rm erg\, s^{-1}\, cm^{-2}$, $F_{opt,\, min} = -11.5\,\rm erg\, s^{-1}\, cm^{-2}$ (values higher than minimum observed).}
    \label{fig:LXLOPT}
\end{figure}

\cite{dainotti2010,dainotti2017,Dainotti2020ApJ...904...97D,Bhardwaj2023} proved that apart from the traditional division of GRBs into Long (LGRBs) and Short (SGRBs), it is vital to consider additional subclasses to constrain well parameters of the 3D Dainotti relation. The authors demonstrated clear differences in the behaviour of LGRBs with and without Supernovae association, as well as in the behaviour of Short GRBs with and without Kilonove association. Thus, in our work, we divide the sample into X-ray-rich (XRR) \citep{Lamb2004} GRBs, X-ray Flashes (XRFs) \citep{Lamb2004}, SGRBs, GRBs associated with SNe (SNe), and LGRBs. We present the plot of $log_{10}(L_X)$ - $log_{10}(L_{opt})$ with division of data into subclasses in Fig. \ref{fig:LXLOPT}. 

\begin{figure*}[ht!]
    \centering
    \includegraphics[width=0.99\linewidth]{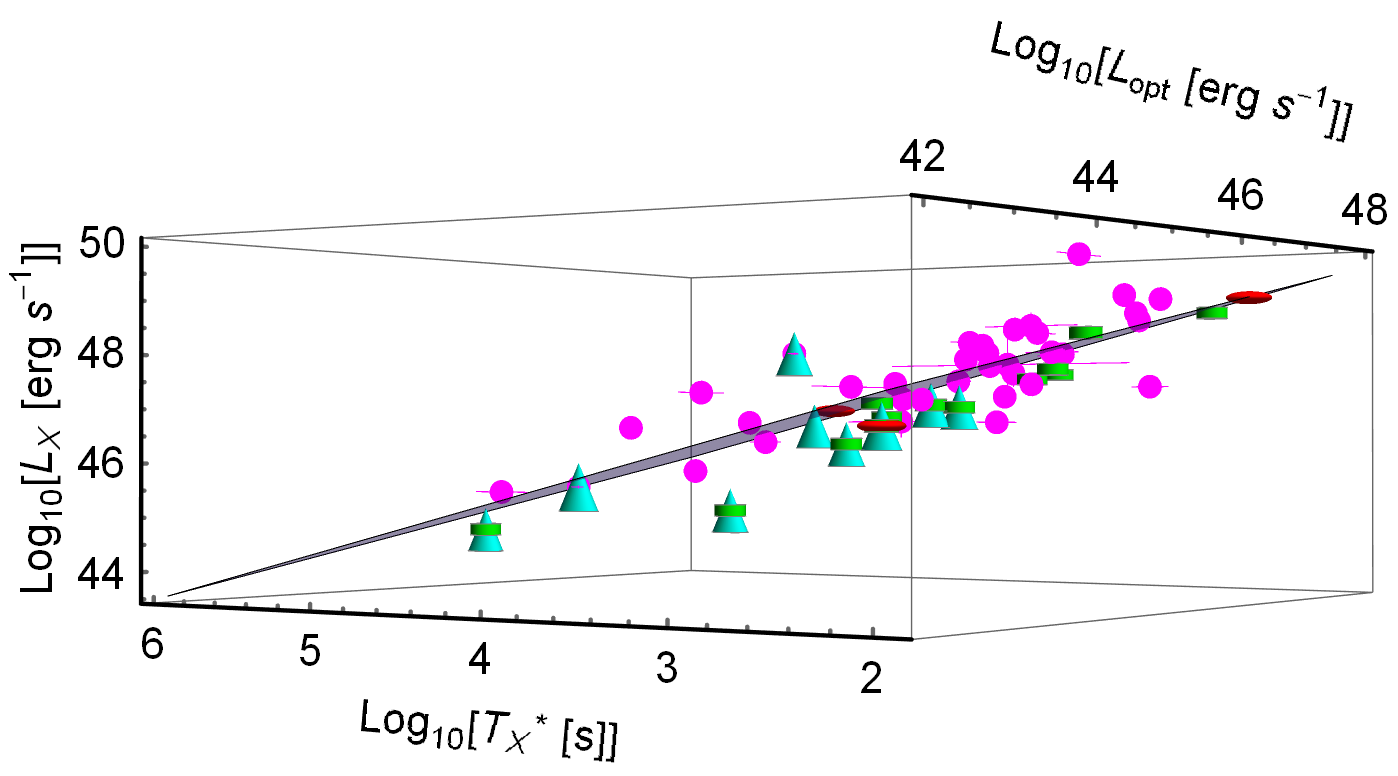}
    \caption{A side on view on the $\log_{10}(L_X)$-$\log_{10}(T^*_X)$-$\log_{10}(L_{opt})$ correlation. Magenta balls represent LGRBs, green cylinders - GRBs identified as XRFs and XRRs, red ellipsoids - SGRBs, and cyan cones - GRBs associated with SNe.}
    \label{fig:LXTXLOPT}
\end{figure*}

No particular clustering of the data can be observed, however, SNe GRBs are present only at $L_{opt}<10^{45.6}\rm \, erg\, s^{-1}$. The correlation seems to be not affected significantly by the Malmquist bias (the minimum observed luminosities are marked by a blue dashed line on the aforementioned plot). The correlation seems prominent, although the scatter is relatively high. Due to the existence of $L_{X}$ - $T_{X}$ correlation, one can expect that the scatter is a result of the projection of a higher dimension correlation. Thus, we investigated different combinations of parameters. The most significant correlation was obtained in the case of $L_X$ - $T_X$ - $L_{opt}$ correlation. 
We present a plot of this new 3D relation in Fig. \ref{fig:LXTXLOPT}. 
The scatter of $log_{10}(L_X)$ - $log_{10}(T^*_X)$ correlation obtained for the studied sample of 52 GRBs is $\sigma_{2D} = 0.74\pm0.07$. Therefore, we observe that the addition of optical luminosity as a third parameter significantly reduces the scatter ($\sigma_{3D} = 0.59\pm0.07$). The scatter of the 3D correlation is $\sim 20\%$ smaller than the scatter of the 2D correlation. To evaluate if the reduction might be an effect of over-fitting we computed the AIC criterium:

\begin{equation}
    AIC = 2k-2\times log\left(\hat{\mathcal{L}}\right),
\end{equation}

where $k$ is the number of estimated parameters in a given model, and $\hat{\mathcal{L}}$ is the maximized value of the postulated likelihood. Further, we calculated the likelihood as in Sec. \ref{sec:analysis}, and obtained that the 3D correlation is strongly preferred over the 2D one with probability $>99\%$.

The best-fit parameters from the MCMC fitting without correction for evolution to different samples are shown in Tab. \ref{tab:opt}. The results of the fitting with correction for evolution are described in Tab. \ref{tab:optEv}.

\begin{table*}[ht!]
    \centering
    \begin{tabular}{|c||c|c|c|c|c|}
        \hline
        Class & N & a & b & c & $\sigma_{int}$ \\\hline\hline
        All & 52 & $-0.88\pm0.12$ & $0.45\pm0.09$ & $30.1\pm4.3$ & $0.59\pm0.07$ \\\hline
        Long & 36 & $-0.81\pm0.16$ & $0.40\pm0.11$ & $32.1\pm5.4$ & $0.61\pm0.08$ \\\hline
        Long without overlap & 31 & $-0.90\pm0.19$ & $0.46\pm0.14$ & $29.5\pm6.3$ & $0.59\pm0.09$ \\\hline
        SNe & 9 & $-0.88\pm0.51$ & $0.40\pm0.39$ & $32\pm18$ & $1.03\pm0.40$ \\\hline
        XRR & 11 & $-1.46\pm0.20$ & $0.41\pm0.11$ & $33.0\pm5.3$ & $0.29\pm0.10$ \\\hline
        XRR+XRF & 13 & $-1.13\pm0.16$ & $0.44\pm0.10$ & $30.8\pm4.6$ & $0.46\pm0.10$ \\\hline
        All without Short & 49 & $-0.92\pm0.13$ & $0.41\pm0.10$ & $31.7\pm4.7$ & $0.60\pm0.07$ \\\hline
    \end{tabular}
    \caption{The chosen samples and the corresponding best-fit parameters of the equation: $\log_{10}(L_{X}\,{\rm[erg\, s^{-1}]})=a\times \log_{10}(T_{X}\, {\rm[s]})+b\times \log_{10}(L_{opt}\, {\rm[erg\, s^{-1}]})+c$. No correction for evolution was applied. (XRR - X-ray rich, XRF - X-ray flashes)}
    \label{tab:opt}
\end{table*}

\begin{table*}[ht!]
    \centering
    \begin{tabular}{|c||c|c|c|c|c|}
        \hline
        Class & N & $a_{evo}$ & $b_{evo}$ & $c_{evo}$ & $\sigma_{int}$ \\\hline\hline
        All & 52 & $-0.86\pm0.13$ & $0.36\pm0.13$ & $33.9\pm5.7$ & $0.58\pm0.07$ \\\hline
        Long & 36 & $-0.78\pm0.17$ & $0.42\pm0.15$ & $31.0\pm7.0$ & $0.58\pm0.09$ \\\hline
        Long without overlap & 31 & $-0.82\pm0.19$ & $0.40\pm0.17$ & $32.3\pm7.9$ & $0.57\pm0.11$ \\\hline
        SNe & 9 & $-0.89\pm0.51$ & $0.37\pm0.42$ & $33\pm19$ & $1.00\pm0.35$ \\\hline
        XRR & 11 & $-1.47\pm0.27$ & $0.43\pm0.22$ & $32.7\pm9.8$ & $0.31\pm0.14$ \\\hline
        XRR+XRF & 13 & $-1.10\pm0.19$ & $0.41\pm0.18$ & $32.0\pm7.9$ & $0.40\pm0.14$ \\\hline
        All without Short & 49 & $-0.90\pm0.14$ & $0.32\pm0.13$ & $35.8\pm6.0$ & $0.59\pm0.08$ \\\hline
    \end{tabular}
    \caption{The chosen samples and the corresponding best-fit parameters of the equation: $\log_{10}(L_{X}\,{\rm[erg\, s^{-1}]})=a\times \log_{10}(T^*_{X}\, {\rm[s]})+b\times \log_{10}(L_{opt}\, {\rm[erg\, s^{-1}]})+c$. Correction for evolution was applied. The luminosity was corrected with the so-called K-correction. (XRR - X-ray rich, XRF - X-ray flashes)}
    \label{tab:optEv}
\end{table*}

\begin{figure}[ht!]
    \centering
    \includegraphics[width=1\linewidth]{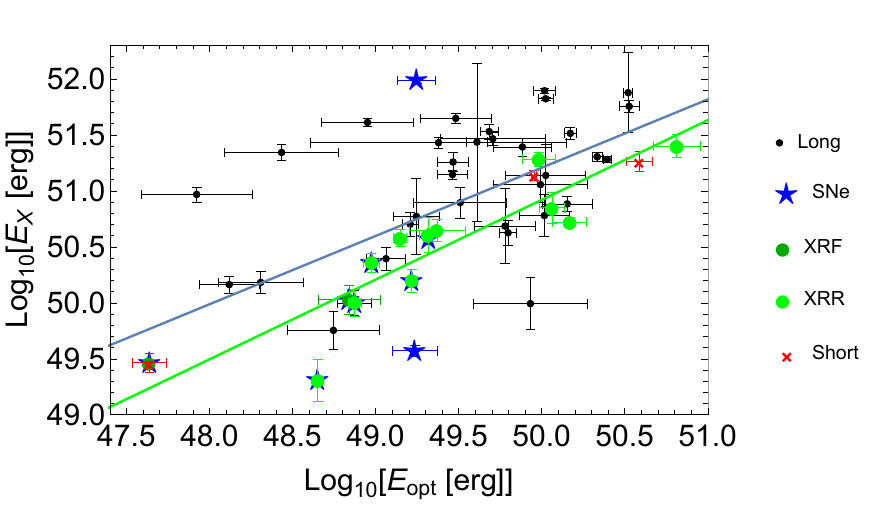}
    \caption{Plot of energy emitted during the plateau measured in X-rays as a function of the corresponding energy measured in optical R band. The style of plot markers distinguishes the class of a given GRB. Long GRBs are presented as black circles, short as red crosses, SNe associated as blue stars, and XRRs together with XRFs are marked as green circles. The blue solid line is a simple linear regression that best fits all data, while the green solid line is a simple linear regression fit only to XRR GRBs.}
    \label{fig:EXEOA}
\end{figure}

The $E_X-E_{opt}$ diagram is shown in Fig. \ref{fig:EXEOA}. Again, no clear difference in data distribution can be seen between classes. Such behaviour may indicate the independence of correlation from the engine and environment type, directing attention to the shock physics. 

\section{The LC with energy injection from the central engine}
\label{sec:BH}

\cite{Dainotti2021PASJ} demonstrated that for a large sample of GRBs, the closure relationships show the preference of the data towards the energy injection scenario in opposition to the scenario of singular jet ejection and gradual cool-down. In other words, the GRB LCs demand that energy is constantly delivered to the shocks, which reprocess this energy and re-emit it in X-rays and optical through a synchrotron emission. The two most common choices of an engine that can realize such a scenario are BHs and magnetars. In the magnetar case, the high magnetic field separates the charges around the neutron star and creates a multipole charge distribution on the surface of a spinning NS, therefore an electromagnetic emission arises \citep{DaiLu1998A&A,DaiLu1998MNRAS,DaiLu1998PhRvL,zhangmeszaros2001,Wang2024}. In principle, both engines are capable of accretion of fall-back material, which delivers energy to the system, providing a good description of observed afterglow with features like bumps and flares \citep{2021ApJ...918...12F}.

\subsection{Millisecond magnetar}

\cite{zhangmeszaros2001} was one of the earliest papers to make predictions of the X-ray plateaus. After the {\it Swift} discovery, the magnetar model was discussed by \cite{Zhang2006} comprehensively. Smoking gun signatures for the magnetar model have been presented by \cite{Troja2007,Lyons2010,Rowlinson2010,Rowlinson2013,Rowlinson2014, LuZhang2014, Rea2015,Lu2015, stratta2018}. 
 Therefore, a rapid-spinning magnetized NS named ``millisecond magnetars" is a likely candidate for some GRB engines. The total rotational energy of a millisecond magnetar is denoted by
\be\label{Erot}
E_{\rm}=\frac12 I\, \Omega^2\,\approx 2.6 \times 10^{52}\,{\rm erg}\,M^{\frac32}_{\rm ns,1.4}\,P^{-2}_{-3}\,,
\ee
where $I\simeq 1.3\times 10^{45}\,M^{\frac32}_{\rm ns,1.4}\,{\rm g\,cm^2}$ \citep{2005ApJ...629..979L} with $M_{\rm ns}=1.4\, M_\odot$ and $R_{\rm ns}= 1.2\times 10^6\, {\rm cm}$  the NS mass and radius, respectively, and $P=2\pi/\Omega$ the spin period. The NS may experience fall-back accretion once the millisecond magnetar has formed \citep{1989ApJ...346..847C, 2007MNRAS.376L..48R}.  The fall-back accretion rate can be described, following \cite{2018ApJ...857...95M}, as:

\be\label{dotM}
\dot{M}=\frac23\frac{M_{\rm fb}}{t_{\rm fb}}
\begin{cases}
1  \hspace{1.6cm}{\rm for} \hspace{0.4cm} t< t_{\rm fb}   \cr
\left(\frac{t}{t_{\rm fb}}\right)^{-\frac{5}{3}} \hspace{0.7cm}{\rm for} \hspace{0.3cm} t_{\rm fb}< t\,,\cr
\end{cases}
\ee

where  $M_{\rm fb}$ and $t_{\rm fb}$ are the accreting mass and the characteristic fall-back time, respectively. This accretion is influenced by the dipole magnetic moment ($\mu=BR^3_{\rm ns}$, with $B$ being the strength of the dipole magnetic field), and the Alf\'en ($r_{\rm m}$), the co-rotation ($r_{\rm c}$) and cylinder ($r_{\rm lc}$) radii. Due to the accretion, the spin evolution can be written as, following \cite{2011ApJ...736..108P}:

\be\label{dif_eq}
I\frac{d\Omega}{dt}=-N_{\rm dip} + N_{\rm acc}\,,
\ee

where $N_{\rm dip}$ and $N_{\rm acc}$ are defined by:

\be\label{N_dip}
N_{\rm dip} \simeq  \mu^2\Omega^3
\begin{cases} 
 \frac{r^2_{\rm lc}}{r^2_{\rm m}}  \hspace{0.8cm} r_{\rm m} \lesssim  r_{\rm lc}\,, \cr
1  \hspace{1.05cm} r_{\rm lc} \lesssim r_{\rm m}\,, \cr
\end{cases}
\ee
and 
\be\label{Nacc}
N_{\rm acc}=\dot{M}(G_N\,M_{\rm ns}\,r_{\rm m})^\frac12\, \left[ 1-\left( \frac{r_{\rm m}}{r_{\rm c}} \right)^\frac32 \right]\,,
\ee
respectively.  Here, $G_N$ is the gravitational constant. The entire evolution of the spin-down luminosity, $L_{sd}$, due to the fall-back accretion is explicitly derived in \citet{2018ApJ...857...95M}. An analytic solution for the spin-down luminosity, once equilibrium is reached, can be written, following \cite{2018ApJ...857...95M,2021ApJ...918...12F}, as:
{\small
\be\label{L_sd}
L_{\rm sd}\approx 10^{40.7}\,{\rm erg\, s^{-1}}B^{-\frac67}_{16}\,M^{\frac{12}{7}}_{\rm ns,1.4}\,R^{-\frac{18}{7}}_{\rm ns,6.1}\,
\begin{cases} 
t^0\,  \hspace{0.6cm}{\rm for} \hspace{0.1cm} t\ll t_{\rm fb}   \cr
 t^{-\frac{50}{21}}_{8} \hspace{0.25cm}{\rm for} \hspace{0.1cm} t\gg t_{\rm fb}\,,\cr
\end{cases}
\ee
}
where the typical values of the accreting mass $M_{\rm fb}=0.8M_{\odot}$ and the characteristic fall-back time $t_{\rm fb}=10^8\,{\rm s}$ are used.  It is worth noting that, the spin-down luminosity before the characteristic fall-back time evolves as $L_{\rm sd}\propto t^{0}$.

\begin{figure*}[ht!]
    \centering
    \includegraphics[width=0.33\linewidth]{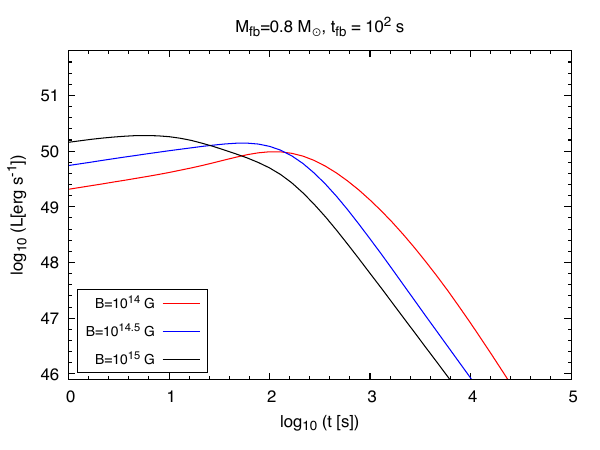}
    \includegraphics[width=0.33\linewidth]{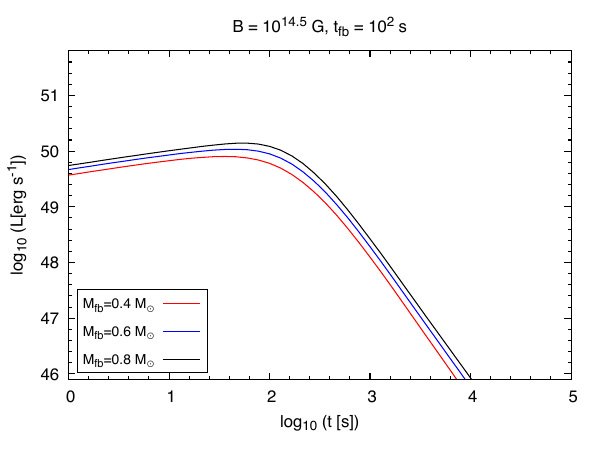}
    \includegraphics[width=0.33\linewidth]{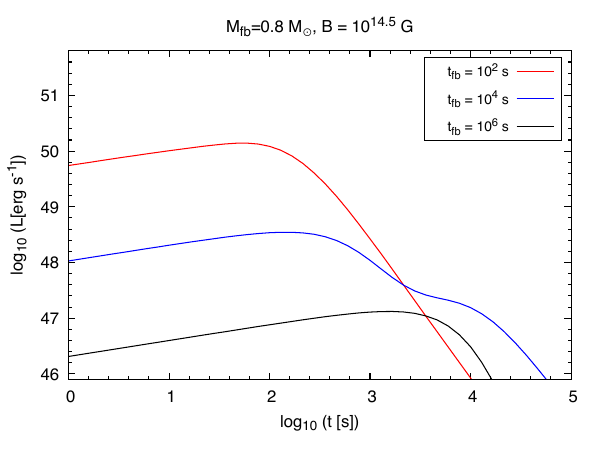}
    \caption{Left panel: the theoretical LCs computed with the millisecond magnetar model with fall-back accretion with initial parameters $M_{\rm fb}=0.4 M_{\odot}$, $t_{fb}=10^2$ s, and $B \in \left\{\;10^{14},10^{14.5},10^{15} \right\}\text{ G}$. Central panel: the same as left, but we fix $B = 10^{14.5} \text{G}$, and vary accretion mass: $M_{\rm fb} \in \left\{0.4, 0.6, 0.8 \right\} M_{\odot}$. Right panel: the same as the other two, but we fix $B=10^{14.5}\,\rm G$, $M_{\rm fb}=0.8 M_{\odot}$, and vary $t_{\rm fb} \in \left\{10^2,10^4,10^6\right\}$ s.}
    \label{fig:MAGthLC}
\end{figure*}

Figure \ref{fig:MAGthLC} shows the evolution of the spin-down luminosity of a millisecond magnetar when accreting for a range of parameter values. The light curves in the left panel are for the accreting mass $M_{\rm fb}=0.8\,{M_\odot}$, the characteristic timescale $t_{\rm fb}=10^{2}\, {\rm s}$ and the strengths of the magnetic fields $B=10^{14}$ (red), $10^{14.5}$ (blue) and $10^{15}\,{\rm G}$ (black),  in the central panel for $B=10^{14.5}\,{\rm G}$, $t_{\rm fb}=10^{2}\, {\rm s}$, and $M_{\rm fb}=0.4$ (red), $0.6$ (blue) and $0.8\,{M_\odot}$ (black), and in the right panel for $B=10^{14.5}\,{\rm G}$, $M_{\rm fb}=0.8\,{M_\odot}$, and $t_{\rm fb}=10^{2}$ (red), $10^{4}$ (blue), and $10^{6}\, {\rm s}$ (black).
The light curves in all panels display plateau phases with different durations and levels depending on the parameter configurations. For instance, as the magnetic field increases, the plateau phase is shorter and the luminosity is higher. As expected, the luminosity is higher and the duration is longer when the accreting mass is bigger.

This model, however, sometimes requires an enormous magnetic field ($B\sim10^{17}\, \rm G$ \citep{Rowlinson2014}) to explain the brightest burst. This estimate is compatible with the super-magnetars scenario suggested by \cite{Rea2015}. The authors analysed a population of galactic magnetars to estimate the maximum magnetic field that can emerge around newborn magnetar and obtained $B\sim 10^{16}\,\rm G$. However, due to the large uncertainties on several parameters, such as the population synthesis models, the hypothesis of the existence of newly born magnetars with $B\sim 10^{17}\,\rm G$ is not rejected, leaving the question open. Despite its issues, the spin-down scenario was successful in predicting the $log_{10}(L_X)$ - $log_{10}(T^{*}_{X})$ anti-correlation \citep{Rowlinson2014, Wang2022ApJ...924...97W}. This suggests that accretion should play a secondary role in most of the bursts, and is needed primarily to explain structures like bumps and flares that live on the spin-down background. This explains, why the removal of such sources from the dataset of LGRBs reduces the scatter of the Dainotti correlation \citep{Dainotti2020ApJ...904...97D}.
However, the $log_{10}(L_X)$ - $log_{10}(L_{\rm peak})$ correlation origin remains to be further investigated. Another possibility is explained by \cite{Hascoet2014} who presented a possible origin of the connection between prompt and plateau phase within the context of microphysical parameter variation. This scenario is likely to be possible only for a fraction of all GRBs. \cite{vaneerten2014a} presented a simple model for shock propagation in the case of injected energy and non-uniform medium. The author derived analytically the functional form of the LC and provided a theoretical explanation of the $L_{X}$ - $T^*_{X}$ and  $L_{opt}$ - $T^*_{opt}$ correlations. In this model, the energy injection from the central engine stops rapidly and the behaviour of the LC after this point is a function only of the total injected energy. A more detailed study is presented in \cite{vaneerten2014b}, where the author studies two cases: thick and thin shells. In the first case the LC behaviour is strongly connected with the energy injected from the central engine as a function of time, while in the latter case, the evolution of shock is constrained primarily by its initial energy during ejection. The author concluded that the existence of the $L_{X}$ - $T^*_{X}$ and  $L_{opt}$ - $T^*_{opt}$ correlations favours the first model. Moreover, in many cases, the energy reservoir of magnetar. Most extreme unstable case entails $1-2\times 10^{53}$ erg \citep{Metzger2015,dallosso2018}) which seems to be an order of magnitude too small to explain the total energy of GRBs \citep{Li2018b}. However, the BH progenitor has been widely studied in the context of prompt emission \citep{Liu_2018,Li2024}. The latter work presents the hyperaccretion scenario, where prompt emission arises when the BH is actively accreting mass while emitting electromagnetic energy through the BZ process. 

\subsection{Afterglow generated by accretion or fall-back}

\cite{Kumar2008} investigate the dynamics of mass fall-back and accretion during the collapse of rapidly rotating massive stars. The authors model the infall rate of stellar material onto an accretion disk surrounding a nascent black hole. Their findings indicate that the resulting relativistic jet maintains a high luminosity ($\sim10^{52}$ erg/s) for approximately 100 seconds (powering the prompt phase). Subsequently, the jet's luminosity diminishes rapidly over the next $\sim10^3$ s. The study addresses the plateau phase observed in the X-ray light curves of certain GRBs, which can persist for $\sim10^4$ s post-burst. The authors propose that these plateaus result from continued accretion, either due to: a low viscosity parameter within the accretion disk, leading to sustained mass accretion rates, or ongoing fall-back of material from an extended stellar envelope or supernova ejecta that failed to achieve escape velocity. The duration and shape of the plateau are controlled by the mass fall-back rate and the distribution of specific angular momentum in the accretion disc. 
\cite{Cannizzo2011} discusses the $L_X$ - $T^*_X$ anti-correlation in context of a similar model \citep{Cannizzo2009}. Their research indicates that the existence of this relation requires an approximately constant amount of total accreted matter across the sources and allows for a correlation slope of -1.2. However, it is still unclear the process which would ensure that the total amount of accreted matter is always the same. More recently \cite{Zhao2025} analysed a sample of 103 sources and successfully explained the plateau phase for some of them via the model of fall-back accretion onto the BH.
It is possible to show that, the BH with $a\approx0$ created during initial collapse will increase its mass and spin during the accretion. As presented by \cite{Li2024,Lei_2017} the hyperaccretion can spin up the BH to $a>0.9$. This leads to a BH with a huge kinetic energy reservoir. Similar results were also obtained by \cite{Shibata2025} with a more theoretical approach, which also included the analysis of the formation of the magnetic field. The initial rapid growth of BH is followed by the accretion of less dense matter from the star's envelope and fall-back material. 
If the created accretion disk has a sufficient magnetic field to enter the MAD regime, the BH rotational energy can be extracted via BZ-driven spin-down. The presence of MAD in GRBs was previously proposed for the prompt phase by, for example, \cite{Gottlieb_2023} for the bursts created in binary mergers and \cite{Gottlieb_2023b,Issa2025} for collapsars. However, recently, \cite{Wu2025} discussed that such a mechanism is not efficient enough to explain the observed prompt luminosities of most GRBs.

\subsection{Black Hole spin-down}
If the BH can be a central engine of GRBs, then the BZ process requires a high magnetic field, which in turn alters the accretion rate. Thus, during the prompt phase, the creation of the BH should be held back by the magnetic field needed to explain the high luminosity of the bursts. However, recently \cite{Gottlieb_2024} proved that thanks to the magnetic field inherited by a BH from the progenitor star it is possible for the BH to launch highly luminous jets. Due to the high efficiency of the \cite{BlandfordZnajek} process, some authors already discussed that the spin of a quickly rotating BH plays a crucial role in powering the plateau \citep{KumarZhang2015}.  More recently \cite{Lei_2017} discussed comprehensively the BH as a central engine of both the prompt and the afterglow phase. The authors concluded that the GRB lightcurve is a result of an interplay between the spin-up process via accretion and the spin-down process via BZ energy extraction. Nevertheless, the authors assumed that the magnetic field is generated by the accretion disk but does not influence its dynamics. Therefore, the accretion is very efficient, and the BH reaches a high equilibrium spin of $a\approx 0.87$. This results in an accretion-driven shape of the lightcurve (the luminosity gets smaller as the engine slowly runs out of the material in the accretion disk), which does not show a characteristic break often associated with the end of the plateau. As we show further in the text, such a break is produced when the BH spin-down is more efficient and the BH reaches a low equilibrium spin of $a<0.1$. In this regime, the BZ process becomes inefficient and luminosity drops rapidly.
\cite{Li2018b} discusses that both magnetar and a BH can drive the plateau via a spin-down process, however, the authors do not argue that this process is responsible for the break in the afterglow but rather present an empirical, qualitative discussion in the context of the energy budget. Interestingly, the authors found that the sample of GRBs with too high ($>2\times 10^{52}$ erg) an energy budget to be driven by a magnetar, follows the $L_X$-$T^*_{X}$ correlation tightly. This encourages us to further study the BH model. The spin-down of both magnetars and BHs was also discussed by \cite{vanPutten2004, vanPutten2023} in the context of gravitational wave (GW) emission and the energy budget, however not in the context of a primary process behind the shape of the afterglow.

We take inspiration from those works and consider the BH in its later stage of evolution (after the rapid accretion driving the prompt emission). BH is quickly spinning $a\approx1$, and the accretion rate is driven primarily by the matter from the envelope of the progenitor star (which was much less dense than the star's core; thus, the accretion rate is much smaller than during the prompt phase). The disk is balanced by the magnetic field, creating the so-called Magnetically Arrested Disk (MAD), and BH emits light in a spin-down scenario powered by a Blandford - Znajek process. In this scenario the equality of magnetic field pressure and pressure due to accretion leads to an approximate formula for the magnetic field $B$, as a function of spin $a$, mass accretion rate $\dot{m}=\frac{\dot{M}}{M_{\odot}s^{-1}}$, and the BH mass $m_{\rm BH}=\frac{M_{\rm BH}}{M_{\odot}}$ \citep{Lei_2017}:

\begin{equation}
    B\approx 7.4\times 10^{16} \frac{\sqrt{\dot{m}}}{m_{\rm BH}} \times \left(1+\sqrt{1-a^2}\right)^{-1}\rm G.
    \label{eq:Bfield}
\end{equation}

Then, the evolution of BH is described by the energy and angular momentum conservation laws:

\begin{equation}
\begin{array}{ll}
    \frac{d M_{\rm BH}}{d t} = \dot{M}e_{in}-\frac{L_{\rm BZ}}{c^2},\\\\
    \frac{d a}{d t} = \frac{c}{G M^{2}_{\rm BH}}\left(\dot{M}l_{in} - T_{B} \right) - \frac{2a}{M_{\rm BH}}\times \left(\dot{M}e_{in} - \frac{L_{\rm BZ}}{c^2} \right).
\end{array}
\label{Econs}
\end{equation}

The above equations say that BH loses mass due to a jetted energy emission via the BZ process with power $L_{\rm BZ}$, while the accretion delivers new energy. The rate of energy increase is controlled by the accretion rate $\dot{M}$ and the specific energy of particles, $e_{in}$, entering the BH  \citep{Bardeen1970}. A similar behaviour can be observed in the spin evolution equation. The angular momentum is delivered by accretion with a rate dependent on the specific angular momentum of particles, $l_{in}$, entering the BH  \citep{Bardeen1970} and is extracted by the magnetic torque $T_B$. The second term in the spin-evolution equation describes the angular momentum change due to BH mass change. The luminosity $L_{\rm BZ}$ of a quickly spinning BH ($0<a<1$) is given by a general formula described in \cite{BlandfordZnajek,Lee_2000,LEE200083,Liu_2018}:

\begin{equation}
\begin{array}{ll}
    L_{\rm BZ} = 1.7\times 10^{20} \left(\frac{B}{1\text{ G}}\right)^2 \left(\frac{M}{M_{\odot}}\right)^2 a^2 F(a) \text{ erg s}^{-1},\\
    F(a) = \left(\frac{1+q(a)^2}{q(a)^2}\right)((q(a)+1/q(a))\arctan(q(a))-1),\\
    q(a)=\frac{a}{1+\sqrt{1-a^2}}.
\end{array}
\label{BZ}
\end{equation}

In the case of Keplerian accretion, $l_{in}$ and $e_{in}$ are the specific angular momentum and energy of particles at the marginally stable orbit, which are approximately constant further inside beyond the horizon. However, in the case of MAD accretion disks, the magnetic field causes the disk to lose angular momentum and energy. This allows for a smaller contribution of accretion to the spin evolution, and extraction via the BZ process significantly dominates. \cite{Narayan2022} showed with numerical simulation, that such spin-down from $a=1$ results in an equilibrium state with $a_{eq}\approx 0.04$. \cite{Lowell_2024} obtained a slightly larger value for modified disk properties. Both values are very low, which indicates an efficient energy loss. 

\cite{Lowell_2024} demonstrated that within the MAD scenario, the spin evolution can be described as:

\begin{equation}
    a(t) = a_{eq} + |a_0-a_{eq}|e^{-t/\tau},
\end{equation}

where $a_{eq}$ is the minimum spin that can be obtained in the considered process, $a_0$ is the initial spin at $t=0$, while 

\be
\frac{t}{\tau}=9.5\frac{\Delta M}{M(0)}
\label{eq:timescale}
\ee
controls the spin-down timescale of accretion in the MAD scenario.

\begin{figure*}[ht!]
    \centering
    \includegraphics[width=0.33\linewidth]{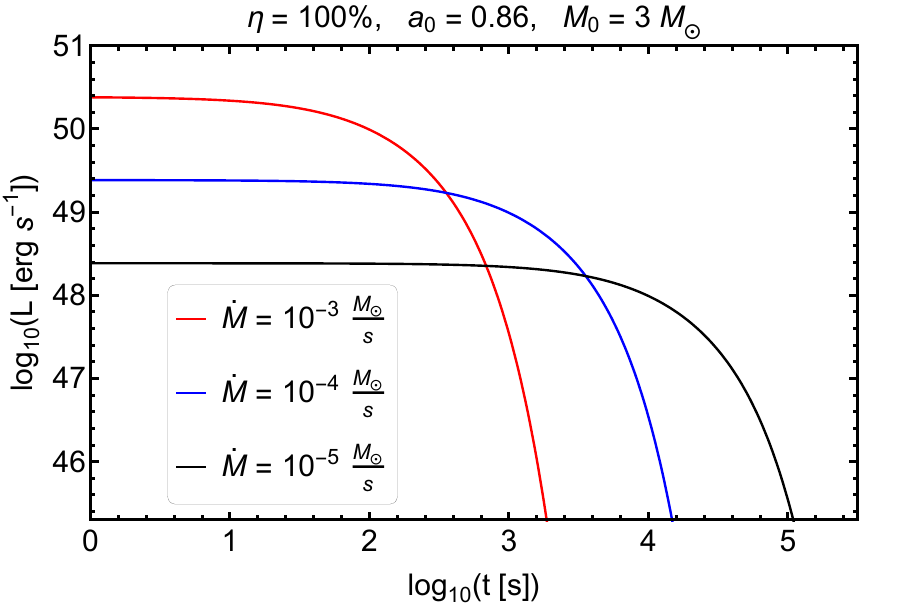}
    \includegraphics[width=0.33\linewidth]{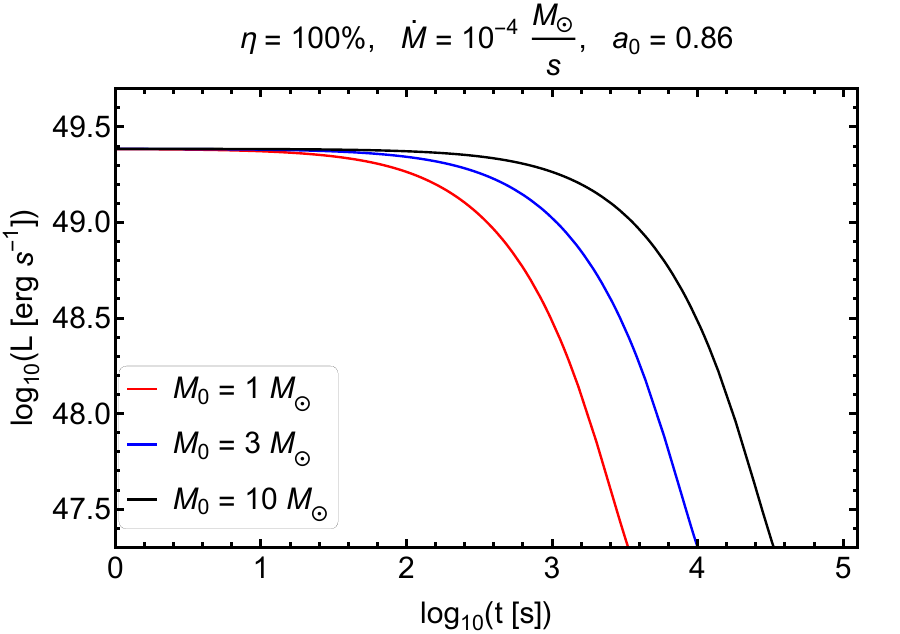}
    \includegraphics[width=0.33\linewidth]{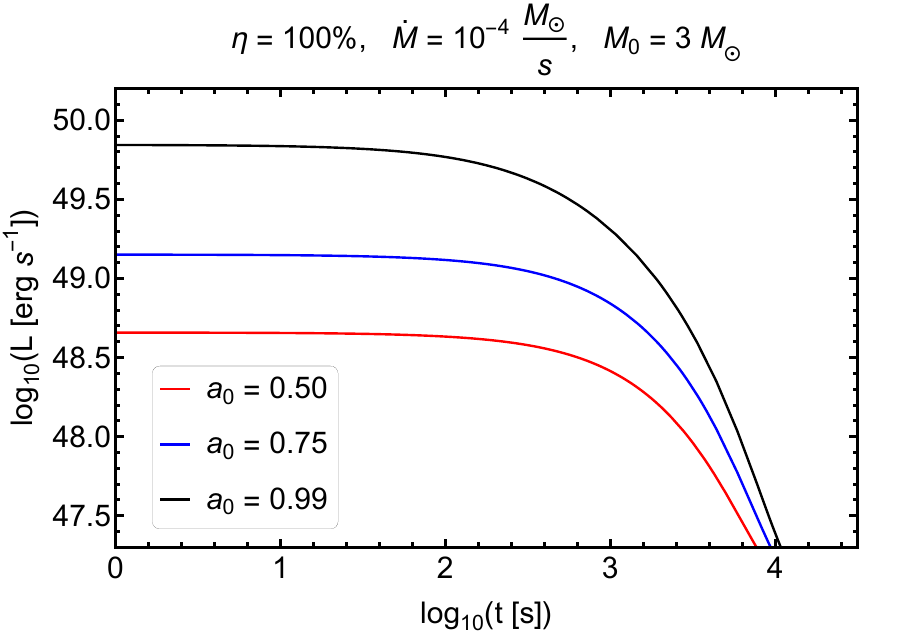}
    \caption{Left panel: the theoretical LCs computed with the Kerr BH spin-down model for BH with initial parameters $M_0=3 M_{\odot}$, $a_0=0.86$, and $\dot{M} \in \left\{\;10^{-3},10^{-4},10^{-5} \right\}\frac{M_{\odot}}{s}$. Central panel: the same as left, but we fix $\dot{M} = 10^{-4} \frac{M_{\odot}}{s}$, and vary mass: $M_0 \in \left\{1,3,10 \right\} M_{\odot}$. Right panel: the same as the other two, but we fix $\dot{M}=10^{-4}\,\rm \frac{M_{\odot}}{s}$, $M_0=3M_{\odot}$, and vary $a_0 \in \left\{0.50,0.75,0.99\right\}$.}
    \label{fig:BHthLC}
\end{figure*}

We present a set of numerical solutions of the time-dependent luminosity for $M(0)\equiv M_0=3M_{\odot}$, $a_{0}=0.86$ (maximum achievable spin via the Keplerian accretion), and $\dot{M} \in \left\{\;10^{-3},10^{-4},10^{-5} \right\}\frac{M_{\odot}}{s}$ see the left panel of Fig. \ref{fig:BHthLC}.

We stress again that the results obtained in this analysis were obtained under the assumption of a nearly constant accretion rate with a duration bigger than the duration of the plateau. However, we also analysed the accretion rate as a function of time with the form $\dot{m}\propto t^\alpha$, with $\alpha\in [-0.5,0.5]$ and still obtained the characteristic break in the afterglow light curve for the range of values of parameters considered in this work. This modification resulted primarily in the change of the slope of the plateau.

\cite{Lowell2025} obtained that a luminous MAD state results in a BH spin-down to a higher value of equilibrium spin ($a_{eq}\sim 0.3$). Such a scenario could be a possible explanation of a shallow-steep-shallow shape of some GRB LCs (e.g. GRB 120213A).

\subsection{The $L_X$ - $T_X$ anti-correlation}
\label{sec:lxtx}
\begin{figure}[ht!]
    \centering
    \includegraphics[width=0.99\linewidth]{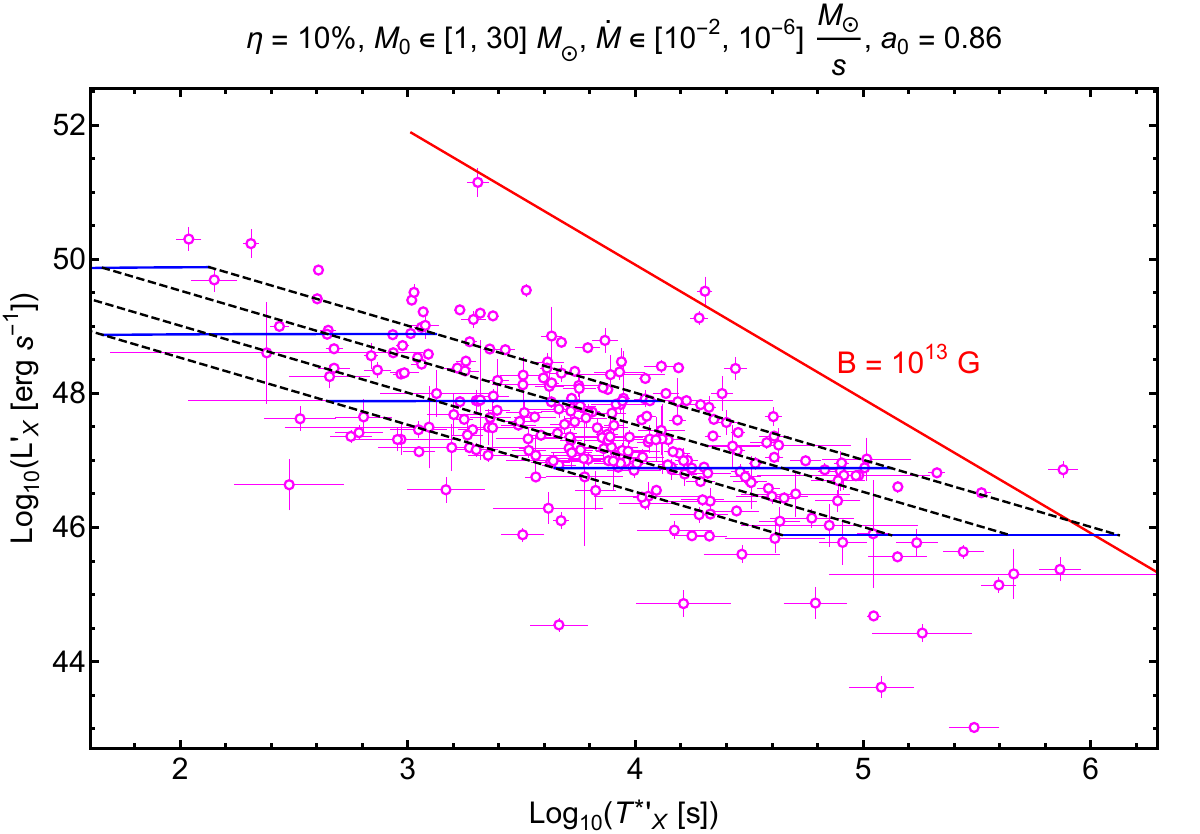}
    \caption{The theoretical region (blue and dashed black lines) in the $L_X$ - $T_X$ space occupied by sources powered by a BH with $a_0=0.86$, $\eta=10\%$, $M_0\in \{1, 3, 10, 30\}\, M_{\odot}$, $\dot{M}\in \{10^{-2}, 10^{-3}, 10^{-4}, 10^{-5}, 10^{-6}\}$. The blue lines were computed for a constant accretion rate each. The lines closer to the top correspond to higher values of $\dot{M}$. The dashed black lines were computed for a constant BH mass each. The higher the line, the more massive the BH. The red line was computed for constant magnetic field $B = 10^{13}\, \rm G$ and efficiency $\eta=100\%$. This line marks the forbidden region, where the magnetic field is too weak to allow for the vacuum polarisation and, therefore, effective production of luminous jets. Sources on the right side of this red line cannot be explained within our simple model.}
    \label{fig:thCor}
\end{figure}

The timescale of a spin-down is controlled by the $\tau=\frac{M_0}{9.5\dot{M}}$ parameter. For the fixed mass of the BH, the timescale is driven primarily either by the magnetic field or accretion rate. Those two factors are coupled with each other, but it is more likely that the accretion rate is controlled by the magnetic field than the other way around. The initial magnetic field could be inherited from the progenitor star and have a random geometry, which affects the accretion rate with a different strength. Therefore, we do further calculations assuming that the magnetic field is the primary cause of the timescale. Taking the derivative of Eq. \ref{eq:timescale} over $t$, and substituting Eq. \ref{eq:Bfield}, it can be easily shown that the $\tau$ scales like: $\tau\sim B^{-2}M_{0}^{-1}$. If the initial masses of BHs created in collapsars do not vary much between the bursts, a given magnetic field controls the timescale of the energy extraction.
On the other hand, the luminosity scales as $L_{\rm BZ}\sim B^2 M_{0}^2$. The assumption of a random (not correlated with other parameters) and sufficiently narrow mass distribution of BH and the distribution of $a_0$, lead to the immediate observation: $L_{\rm BZ}\sim \tau^{-1}$. Such criteria lead automatically to luminosity-time anti-correlation driven primarily by the magnetic field. It is important here to note that an almost constant and long-lived accretion rate is necessary to sustain a high magnetic field. However, such conditions are likely to be met by some LGRBs due to the structure of high-mass progenitor stars \citep{Kumar2008}, which have a dense compact core and less dense, extended envelope.
In practice, we cannot directly measure the spin of the black hole and the magnetic field or accretion rate. Thus, the empirical time scale of the plateau $T$ is usually determined by fitting the light curve. Empirically, we observe that the time of the plateau's end is the time when luminosity drops $\sim 0.5\,\rm dex$ compared to the modelled plateau luminosity at t=0. Thus, for further calculations, we define: $log_{10}(L_{\rm BZ}(t=0)) - 0.5 = log_{10}(L_{\rm BZ}(T))$, that allows us to compute the theoretical time of the end of plateau $T$. We highlight that, most likely, the BHs powering GRBs do not have the same mass and initial spin. This introduces scatter to the studied correlation. We show the theoretical region that should be populated by sources with $a_0=0.86$ (maximum achievable spin via Keplerian accretion)  in Fig. \ref{fig:thCor}. Those results were obtained under the assumption of constant efficiency $\eta = 10\%$. This factor is a result of the energy dissipation in shock, jet collimation, and its inclination. Those variables are not measurable with this simple model, thus, we employ the effective parameter $\eta$ to compare our theory with data.

We note that the majority of sources fit very well the theoretical region constrained by the chosen values of BH mass $M_0=1-30 M_{\odot}$ and accretion rate $\dot{M}=10^{-2}-10^{-6} \frac{M_{\odot}}{s}$. It is a fairly easy task to manipulate the considered parameters to construct a region populated by all GRBs. 
The effective jet emission requires a magnetic field strong enough to allow a vacuum polarisation \citep{BlandfordZnajek}. Although even a relatively small magnetic field ($B>10^9$) contributes to the pair production, the spontaneous effective vacuum polarisation requires $B>10^{13}G$ \citep{Schwinger1951,Harding2006}. The choice of this value ensures that a stable magnetosphere forms and charges are constantly delivered to jets, where they are accelerated by the magnetic field. It is worth noticing that pairs can also be produced in other physical processes like photon-photon interaction, which may increase the efficiency of jets even in lower magnetic field scenarios. We computed the region restricted by this border and marked it with a red line in Fig. \ref{fig:thCor}. The sources above this line either have too high a luminosity or plateau duration to be easily explained by our model. Those sources could be maximally collimated (resulting in $\eta>100\%$), or the plateau could be the result of a supernova bump. Given the very prominent degeneracy of the theoretical parameters, it is not possible at this stage to constrain the physical parameters of engines of singular sources. However, further studies that model prompt and afterglow emission could reduce or eliminate the degeneracy and provide our estimates, which would be later comparable with the distribution of BHs observed in our galaxy and its neighbourhood. 
In the middle and left panels of Fig. \ref{fig:BHthLC}, we show the influence of the mass and initial spin on the LC. 

We know that the dispersion of this relation is reduced when $L_{\rm peak}$ is introduced as a third parameter. 
Considering the hyper-accretion scenario, \cite{Li2024} showed that during the prompt phase, BH can actively grow, which manifests in highly energetic emission. If the BH starts its growth with $M=3M_{\odot}$ and $a=0.5$, the majority of BHs do not grow above $3.3M_{\odot}$, but the distribution of final spin is very wide. Therefore, we expect the $L_{\rm peak}$ to depend primarily on $a$. The plateau luminosity $L_{\rm BZ}(T)$ depends strongly on $M$ and $a$, while the timescale depends only on $M$. This explains why we observe strong $L_X$ - $T_X$ and $L_X$ - $L_{\rm peak}$  correlations, while the $L_{\rm peak}$ - $T_X$ relation remains weak. Therefore, the fundamental plane correlation is likely to be a natural consequence of the BH spin-down model. \cite{Dainotti2011} demonstrated that apart from known $L_X$ - $L_{peak}$ correlation, there exists also the $L_X$ - $E_{iso}$ correlation. Given that $E_{iso}$ is sometimes considered to be correlated with the kinetic energy absorbed by a BH via hyperaccretion during the prompt phase \cite{Li2024}, we also deduce that the BH spin-down model presents a feasible explanation for the observed relation.

It is important to notice that the scatter is also, to some extent, the result of non-ideal energy reprocessing by shocks. As we indicated in Sec. \ref{sec:LxLo}, the observed data shows traces of a complicated physics driving the shocks, thus, it is doubtful that the efficiency of shocks has a narrow distribution. However, the presented results hold great promise since, with better data quality, we will be able to comprehensively model the multi-wavelength data and give an estimate of the physical properties of shocks. 
As it has already been discussed in \cite{stratta2018} for the magnetar model, it is important to reduce as much as possible the scatter of the relation due to the observational limitations of the data quality, but there is an inherent scatter which is intrinsic and it is due to the natural variation of the physical properties of the model, such as spin, masse, accretion rate etc.
We do our best to ensure the most precise data possible, however, the GRB LC data often has gaps or was observed for a limited amount of time. 
In this regard, some of us are approaching this issue from the point of view of data quality with machine learning analysis, which allows for a detailed LC reconstruction, see \cite{Dainotti2023a, Manchanda2025}.

\subsection{Future studies on distinguishing between magnetar and BH central engine}

Another important point that we would like to highlight is the existence of degeneracies among the two theoretical scenarios, and that in many bursts, both scenarios can be plausible. It is important to stress that in the context of the magnetar scenario, we can, in principle, exclude the cases with $E_{iso} > 10^{53}$ erg and modelled NS period $P<0.56$ ms \citep{dallosso2018}. On the other hand, cases with too high a magnetic field can be either considered as super-magnetar cases \citep{Rea2015}, or they may be instead considered powered by a BH spin down. A future analysis will entail a one-to-one comparison among the sources to investigate more about the singular cases. This will potentially allow us to definitively pinpoint BH driven plateaus.

\subsection{The difference between LGRBs and SGRBs}

The BH mass resulting from an NS-NS merger usually cannot exceed $\sim 3 M_{\odot}$ (since the mass of each NS is around $1.4\, M_{\odot}$). In the most extreme cases, it could be speculated that NS mass could reach $3.2\, M_{\odot}$ \citep{1974PhRvL..32..324R}, but the merger of such NSs would still have mass $<6.4\, M_{\odot}$. However, such NS would not be stable and would likely collapse to a BH during spin-down long before the potential creation of binary with the other NS. On the other hand, it was shown that the mass of BHs resulting from LGRBs can grow up to $\sim 30\, M_{\odot}$ via accretion \citep{Janiuk2023}. 
Therefore, one expects a $\sim 1$ order of magnitude difference between the masses of BHs created in long and short GRBs. Assuming that both SGRBs and LGRBs create a long-lived highly magnetized disk, the luminosity of the plateau scales as $M^{2}_{BH}$ thus, the $\sim 2$ orders of magnitude difference in $L_X$ between LGRBs and SGRBs (visible in Fig. \ref{fig:ResEisoEpeak}) can be easily attributed to the difference in mass of central BH. 
However, this scenario requires that the mass distribution of BHs with LGRB origin is sufficiently narrow and that the dispersion of the $L_X$ - $T^*_X$ correlation is due to varying efficiency and the shock re-emission effects rather than the wide distribution of BH mass. Moreover, it is likely that the environment of the bursts also plays a crucial role in the clustering. SGRBs, present in under-dense regions outside the galaxy disk, can have a smaller energy transfer efficiency from BH to the shock than LGRBs, which rise in dense star-forming regions. 
The mass-driven clustering is also a suitable explanation for the findings of \cite{Dainotti2020ApJ...904...97D}. The authors showed that the sample of 8 GRBs with hints of Kilonove counterparts all fall below the Dainotti relation determined for LGRBs. Moreover, those GRBs follow their own very tight correlation. The small dispersion observed in this case may be attributed to a nearly constant BH mass. 
Another feasible explanation is the magnetar origin of SGRBs. Due to the lack of material that would ensure a long-lasting accretion disk supporting the magnetic field, which would allow the afterglow emission, it is speculated that the NS-NS merger produces a high-mass magnetar with total kinetic energy $\sim10^{53}\,\rm erg$ \citep{Zhang2025}. This model explains well multiple features observed in SGRBs, making it a plausible scenario \citep{Troja2007,Lyons2010,Rowlinson2010,Rowlinson2013}. On the other hand, we can estimate the energy reservoir of rotating BH as a difference between energy obtained for $a=1$ and $a=0$. One can show that BH can lose even $\sim 29\%$ of its initial mass \citep{Christodoulou1971}. Therefore, an energy reserve of a maximally spinning BH with initial mass $30M_{\odot}$ is $\sim 10^{55}\,\rm erg$. Once again, we reproduce the $\sim 2\,\rm dex$ difference between the two cases. 
The distribution of galactic NS-NS binary systems shows that the components usually have small mass $\sim 1.33\,\rm M_{\odot}$ each \citep{Farrow2019}. Thus, it is likely that their merger will result in a high-mass NS rather than a BH. This makes this scenario an appealing explanation. Moreover, not all SGRBs cluster exactly $\sim 2\,\rm dex$ below the Dainotti relation for LGRBs; some are just slightly more than $\sim 1\,\rm dex$ below. Those brighter SGRBs could be produced by a BH - NS merger, enabling higher energy to power the LC.
The discovery of GWs associated with NS-NS merger GRB 170817A \citep{2017ApJ...848L..12A} marked the beginning of a new era in GRB research. In Fig. \ref{fig:DivEpeakEiso}, it seems that this event is an outlier of both regions populated by both collapsars and mergers. However, this is due to the lack of low-luminosity SGRBs in our sample. \cite{Goldstein2017} showed that this is an ordinary SGRB and confirmed that at least some of such bursts originate from NS-NS mergers. One can speculate that future GW observations have the potential to distinguish between the BH and magnetar models. \cite{vanPutten2023} suggests that the precise GW calorimetry can differentiate between the spin-down of a magnetar and BH. Therefore, future GW observations might be a key to testing the described theories and help to remove some degeneracy between parameters present in LC-based modelling. Moreover, they might be a key to model-independent and distance ladder-free calibration of GRBs \citep{WangWang2019}. 

The outlined theoretical interpretation reassures us that the classification based on $EH$ and $PS$ parameters properly determines the origin of individual sources. The described procedure is highly beneficial for the community, given that one needs to simply measure $L_{X}$, $T^*_{X}$, $L_{peak}$, $E_{iso}$, and $E^*_{peak}$, substitute them to Eq. \ref{eq:EH} and Eq. \ref{eq:PS}. Then, one can identify the localization of a given source at the EH-PS diagram (Fig. \ref{fig:EHPS}) to estimate its class.

\subsection{Plateau-less GRBs}

The BH spin-down due to BH-jet interaction in GRBs in the context of prompt emission was a subject of multiple studies \citep{vanPutten2001,vanPutten2001b,vanPutten2003,vanPutten2012,vanPutten2015,vanPutten2024}, and we note that indeed the theoretical framework assumed above in the text is similar to the framework of aforementioned articles. \cite{vanPutten2012} discussed that the spin-down during prompt can generate the drop of luminosity in high energy LCs for BATSE GRBs. However, this model is not a good approximation for all sources. We note that most high-quality Swift GRBs with a clear presence of plateau (Platinum sample compiled by some of us in \cite{Dainotti2020ApJ...904...97D}) do indeed show an increase in luminosity measured by BAT, which rapidly falls right before the start of the plateau. This allows us to speculate that the lack of the plateau in some LGRBs might be a result of spin-down occurring during the prompt phase. Whether the spin extraction takes place during prompt or afterglow phases depends on the magnetic field around the BH. The magnetic field above $10^{16}\,\rm G $ formed rapidly around the BH, will produce a plateau-less GRB.

\subsection{Metallicity driven evolution}
\label{sec:thEvol}
As we pointed out in Sec. \ref{sec:evolution}, the physical properties of GRBs undergo a significant cosmological evolution. This fact makes the cosmological application of GRBs difficult, but the form of evolution is another factor that should be predicted by the theoretical model.
Higher metallicity stars experience stronger stellar winds than lower metallicity ones. Therefore, during their lifetime they more efficiently lose mass, providing a lower matter supply for a growing BH during core collapse and fall-back at the death of the star \citep{Woosley2006ARA,Woosley2012} (see \cite{Volpato2024} for detailed calculations relating the progenitor characteristics with redshift). Moreover, the efficiency of accretion drops with the growth of metallicity, resulting in reduced growth of BH during the GRB prompt phase. Given the observed inverse trend of metallicity with redshift one expects the correlation of LGRB's BH mass with z. Recall that within our model the GRB observable scales as $L_X \sim M^2$, $T^*_X \sim M^{-1}$. Then, if we can model mass evolution as $M(z)=M_0\times g(z)$, one obtains $L_X(z)=L_0\times g^2(z)$ and $T^*_X(z)=T_0\times g^{-1}(z)$. This result is a perfect match of our qualitative measurements from \ref{sec:evolution}. Namely, distributions of luminosity and time at the end of the plateau scale as $L_X(z)=L_0 \times (1+z)^{2.42\pm 0.58}$, $T^*_X(z)=T_0 \times (1+z)^{-1.25\pm 0.28}$, respectively. \cite{Volpato2024} discusses that such mass evolution could result in GRBs luminous enough to be observed even up to $z\approx 20$.
\cite{lamb2000,2023ApJ...947...85L} presented a comprehensive analysis showing that the metallicity of the initial star strongly affects the free-fall timescale, which is associated as a measure of the observational $T^*_{90}$ parameter. The high-z initial stars had very small metallicity, resulting in a small stellar radius and more rapid and effective accretion onto newborn BH. Currently, the GRB engines are formed from high metallicity stars with bigger radii and less effective accretion, thus longer and dimmer bursts are produced. 

\section{Shock re-emission}

\label{sec: shock re-emission}


Increases or decreases in the values of the fraction of energy given to amplify the magnetic field charcterized by$\epsilon_{\rm B}$) and to accelerate electrons characerized by $\epsilon_{\rm e}$ have also been required during the afterglow to explore a different evolution of the standard synchrotron light curves \citep[e.g., see][]{2003ApJ...597..459Y, 2003MNRAS.346..905K, 2006MNRAS.369..197F, 2006A&A...458....7I, 2020ApJ...905..112F}.  For instance, \cite{2006A&A...458....7I} analysed the temporal evolution of the microphysical parameter $\epsilon_e$ to describe the plateau phase exhibited in several X-ray light curves. The authors reported a best-fit value of $\epsilon_e\propto t^{-\alpha_a}$ with $\alpha_a=1/2$. \cite{2006MNRAS.370.1946G} considered the variation of microphysical parameters ($\epsilon_{\rm e}\propto t^{-\rm \alpha_a}$ and $\epsilon_{\rm B} \propto t^{-\alpha_b}$) to model the X-ray afterglow light curves, reporting a relationship between the power indices of $\alpha_{\rm e}+\alpha_{\rm B}\sim1-2$. \cite{2020ApJ...905..112F} required the variation of the microphysical parameters with $\alpha_a= 0.319$ and $\alpha_b= 1.401$ to describe the Sub-GeV gamma-ray and X-ray light curve of GRB 160509A.


The 3D correlation has decreased the scatter compared to the 2D correlation. This raises an important question of why L$\mathrm{_{opt}}$ at the end of the plateau plays an important role along with L$\mathrm{_X}$ at the end of the plateau. \citet{2023A&A...675A.117R} have investigated these plateau luminosities in X-ray and optical bands simultaneously to give us a physical explanation of the plateau emission in the afterglow phase. Their study utilizes a combined spectral analysis using X-ray and optical data during the plateau phase for a sample of 30 GRBs. $\sim$2/3rd of the GRBs (19/30) indicates that the existence of plateau is a result of the synchrotron emission from a single population of shock-accelerated electrons. This may lead to a correlation of L$\mathrm{_X}$ and L$\mathrm{_{opt}}$. The authors and references therein have discussed various reasons in support of single-zone synchrotron emission. In short, we summarize the theoretical models, such as high-latitude emission (HLE) from structured jets or energy injection from a millisecond magnetar from their paper.

For HLE in structured jets, it is expected that photons emitted off-axis to the observer, i.e., at high latitudes, reach at later times (past the prompt emission) and are less Doppler boosted. The HLE photons retain the shape of the prompt emission spectrum. As they are released at later times, they will modify the shape of the afterglow spectrum, resulting in the plateau. Due to a decrease in Doppler boosting at higher latitudes, the spectral peak and the characteristic cooling frequency in the afterglow spectrum will decrease with time. This evolution, combined with the assumption of HLE, suggests that the synchrotron emission is from a single emission site. The temporal evolution of cooling frequency analyzed by \citet{2023A&A...675A.117R} is consistent with previous studies as well. This indicates that the plateau emission emerges due to photons in the deceleration phase of the forward shock when the jet is observed off-axis at high latitudes.

In the case of the energy injection scenario, a single synchrotron spectrum is favoured as the injection of the additional energy from the millisecond magnetar into the forward shock only impacts the evolution of blast wave dynamics. The evolution of blast waves primarily depends on the rate of energy injection and conversion efficiency from injected energy to jet kinetic energy. While the injected energy modifies blast wave dynamics, it does not influence the energy dissipated by the particles. Thus, the SED in both wavelength bands remains unaffected due to this additional energy and leads to a consistent synchrotron emission spectrum from a single emission site.

Furthermore, 11 GRBs ($\sim$ 1/3rd sample) in their sample are not compatible with the single synchrotron spectrum, indicating that the optical and X-ray plateau emission could be due to two/more emission sites. This will cause either optical or X-ray flux to dominate over the other, in turn leading to a different scaling between $\mathrm{L_{opt}}$ and $\mathrm{L_X}$ values. Alternatively, the X-ray and optical flux could be due to different emission mechanisms. To verify which of these cases is dominant, detailed modelling of the broad-band spectrum is required. 

Either of these scenarios could be responsible for the existence of a $\mathrm{L_X}- \mathrm{L_{opt}}$ correlation. When we combine this with the $\mathrm{L_X}- \mathrm{T^{*}_{X}}$ correlation in Sec. \ref{sec:lxtx}, we obtain an idea of the physical reasoning behind the $\mathrm{L_X}-\mathrm{T^*_{X}}- \mathrm{L_{opt}}$ correlation. However, further modelling of the time-resolved spectrum could help break the degeneracy between the single or multiple emission sites scenario in the synchrotron spectrum and reveal the underlying physics. Such time-resolved analysis will be carried out in a future study.




%
%



\section{Summary and Conclusions}
\label{sec:conclusions}

In this study, we analysed the multi-wavelength correlations of GRBs during their prompt and plateau phases, focusing on emissions across gamma, X-ray, and optical bands. Leveraging data from {\it Swift}, {\it Fermi}, and ground-based optical telescopes, we explored key relationships, such as those involving prompt peak luminosity $L{peak}$, plateau luminosity $L_X$ and $L_{opt}$, and plateau duration $T^*_X$ and $T^*_{opt}$. The results of this investigation contribute significantly to a deeper understanding of GRBs and their potential use as standard candles in future cosmological studies. We provide here a concise summary of our findings in bullet points and discuss it further in the following subsection.
\begin{itemize}
    \item Employment of $L_{peak,\, Fermi}$ in the $L_{X}$-$T^*_{X}$-$L_{peak}$ correlation reduces the scatter by $\sim43\%$ compared to the $L_{peak,\, Swift}$.
    \item The $L_{X}$-$T^*_{X}$-$L_{peak}$ correlation, used together with $E_{iso}$-$E^*_{peak}$ correlation, is a powerful tool to distinguish different classes of GRBs. One has to simply calculate residuals with the provided recipe and locate GRB on the $EH$-$PS$ diagram (Fig. \ref{fig:EHPS}).
    \item A new model was proposed to explain the presence of the plateau in the most energetic bursts. The magnetic field from the MAD disc effectively extracts rotational energy from the Kerr BH, causing a spin-down.
    \item We demonstrate a presence of non-linear $L_{X}$-$L_{opt}$ correlation, its extension to $L_{X}$-$T^*_{X}$-$L_{opt}$ and discuss its possible origin.
\end{itemize}

\subsection{Summary of Findings}

Our work identified several crucial correlations and advancements in GRB studies. Specifically, in Sec. \ref{sec:analysis}, we showed that using {\it Fermi}'s GBM data for peak luminosity $L_{\text{peak, Fermi}}$, rather than {\it Swift}'s BAT data $L_{\text{peak, Swift}}$, leads to a substantial reduction in the scatter of the Dainotti relation. This finding, presented in Tab. \ref{tab:Fermi}, highlights a decrease in intrinsic scatter from $\sigma_{\text{int, Swift}} = 0.46 \pm 0.05$ to $\sigma_{\text{int, Fermi}} = 0.32 \pm 0.03$ without application of correction for evolution. Correcting for redshift evolution further improved the scatter to $\sigma_{\text{int, Fermi}} = 0.25 \pm 0.04$ and $\sigma_{\text{int, Swift}} = 0.41 \pm 0.05$, underscoring the importance of high-quality observations. Therefore, we achieved enhanced precision in the Dainotti relation, improving the GRB's reliability as standard candles. This improvement is particularly important for studying the early universe, where GRBs serve as one of the few tools to probe high redshifts. The findings of this study reaffirm the utility of GRBs as cosmological probes. 

The clustering of GRBs in $L_X$ - $T^*_X$ - $L_{\rm peak}$ parameter space, provides insights into their progenitors. Specifically, Fig. \ref{fig:DivEpeakEiso},\ref{fig:ResEisoEpeak} and \ref{fig:EHPS} illustrate that SGRBs have statistically lower $L_X$ when compared to long bursts with the same $T^*_X$ and $L_{\rm peak}$. Our results on clustering of SGRBs and LGRBs concerning the \cite{amati2006} relation have been compared with those in the literature. We follow  \cite{DelVecchio2016}, who presented an analysis for a deviation from the Dainotti correlation and \cite{Minaev2020}, who introduced the energy-hardness ($EH$) parameter to identify outliers of the Amati correlation. Thus, we define the plateau-shift ($PS$) parameter. There is a clear correlation between clustering of sources in the $EH$ and $PS$ parameter space, and belonging to long and short classes. This clustering offers a method to identify GRB origins when direct observations are incomplete. The clustering analysis demonstrated the critical role of parameters like plateau luminosity ($L_X$) and duration ($T^*_X$). The identification of distinct regions in parameter space for collapsars and mergers further enhances our understanding of GRB origins. Such distinctions are essential for refining GRB classifications and for improving the accuracy of cosmological constraints derived from these events.

We introduced a theoretical BH spin-down model to explain the plateau phase. Based on the Blandford-Znajek process, this model describes the energy extraction from a spinning BH surrounded by the magnetic field, assuming the long-lived MAD regime. We demonstrated an analytical derivation of the empirical $L_X - T^*_X$ correlation within the discussed model. Numerical solutions presented in Fig. \ref{fig:thCor} demonstrate this correlation as a function of the physical parameters of the BH engine. We obtain that observational data can be explained within the new proposed model by a population of BHs with masses $M\in \left[1,30\right]M_{\odot}$ surrounded by magnetic field $B\in \left[10^{13},10^{15.2}\right] G$ generated by a MAD accretion disk with accretion rate $\dot{M} \in \left[10^{-2},10^{-6} \right]\, \frac{M_{\odot}}{s}$. 
The model aligns well with observed correlations and highlights the role of black holes as central engines in GRBs. This theoretical framework is successful in explaining the observed differences in values of $L_X$ observed for LGRBs and SGRBs. Moreover, the redshift evolution of $L_X$ and $T^*_{X}$ can be associated with the engine mass evolution driven by different metallicities of progenitor stars at low and high z.

Additionally, we examined correlations involving optical luminosity at the plateau's end, though these results were limited by sample size. We also highlight the non-linear correlation between X-ray and optical luminosities, suggesting avenues for further exploration with larger datasets.

\subsection{Future Perspectives}

Future research should focus on expanding the GRB sample size, particularly for optical observations, to crucially refine the correlations presented here. Enhanced instrumentation and coordinated observations across multiple wavelengths will also help reduce uncertainties and improve data quality. The perspective on the near future is optimistic based on the recent launch of the Space Variable Objects Monitor (SVOM) satellite, with optical follow-up with Ground Wide Angle Cameras (GWACs) and Ground Follow-up Telescopes (GFTs) \citep{SVOMwei2016deeptransientuniversesvom}.
Additionally, the mission objectives planned for THESEUS could enlarge the GRB sample by an order of magnitude.
An estimated impact of our results can reach the full extent when we consider machine learning analysis \citep{Dainotti2024ApJ...967L..30D,Dainotti2024GRBRedshift,Narendra2024}, which more than double the sample of already observed GRBs by employing a redshift estimator and LC reconstruction. As detailed in \cite{Dainotti2022MNRAS.514.1828D} such enlargement of the sample could allow us to make GRBs a stand-alone cosmological probe already with the current set of observations.
Moreover, theoretical models should be further developed to account for the diverse environments and progenitors of GRBs. For instance, incorporating metallicity-driven evolution could provide a deeper understanding of how GRB properties change with redshift. This is a particularly important issue due to the observed discrepancy between the star formation rate and the GRB formation rate (Khatiya et. al in prep). The analysis of high redshift GRBs is crucial to understanding if the super-massive BH could rise from a merger of stellar-mass BHs. The recent observations by JWST, show that the high-z quasars may be too bright to be explained by the current paradigms. Thus, they would require rapid growth in the early universe, which would require a high GRB formation. Furthermore, exploring the interaction between GRB environments and engine properties, such as BH mass and spin, could shed light on the underlying physics of these extraordinary events.

Future studies on individual GRBs and the full observed population can provide a reliable test of the discussed magnetar and BH models, and possibly pinpoint samples of cohesive origin. Thus, providing an enlarged and physically motivated sample of standardised sources feasible for cosmological computation at high redshift, filling the gap between Supernovae Ia and the Cosmic Microwave Background. Such a high-precision set can be a key to resolving cosmological tensions \citep{Dainotti2021ApJ...912..150D} .

In conclusion, this study demonstrates the power of multi-wavelength observations and theoretical modelling in advancing GRB research. By reducing scatter in fundamental correlations and providing a physical basis for observed behaviours, we take a significant step toward realizing the potential of GRBs as standard candles and probes of the early universe.

\section*{Acknowledgements}

This work made use of data supplied by the UK {\it Swift} Science Data Centre at the University of Leicester. NF acknowledges financial support from UNAM-DGAPA-PAPIIT through the grant IN112525.  We thank Sebastian Szybka and Przemysław Podleśny for the helpful discussion on the Kerr BH energy extraction, Biagio de Simone and Aditya Narendra for the helpful discussion on the GRB observations. We thank Stanisław Zoła for the useful comments on the text.

\section*{Data Availability}

This article did not generate new data products. This work made use of data supplied by the UK {\it Swift} Science Data Centre at the University of Leicester. The X-ray {\it Swift} LCs are accessible via online tool \href{https://www.swift.ac.uk/burst_analyser/}{burst analyser}. The optical LCs can be found in the \href{https://grblc-catalog.streamlit.app/}{public repository}. The {\it Swift}-BAT data is available in the \href{https://swift.gsfc.nasa.gov/results/batgrbcat/index_tables.html}{NASA catalogue}. The {\it Fermi} GBM data is accessible in the online \href{https://heasarc.gsfc.nasa.gov/w3browse/fermi/fermigbrst.html}{repository} \citep{Gruber_2014,vonKienlin_2014,Bhat_2016,vonKienlin_2020}.

\appendix

\section{The Efron \& Petrosian method}
\label{Appendix}

\begin{figure}[ht!]
    \centering
    \includegraphics[width=0.99\linewidth]{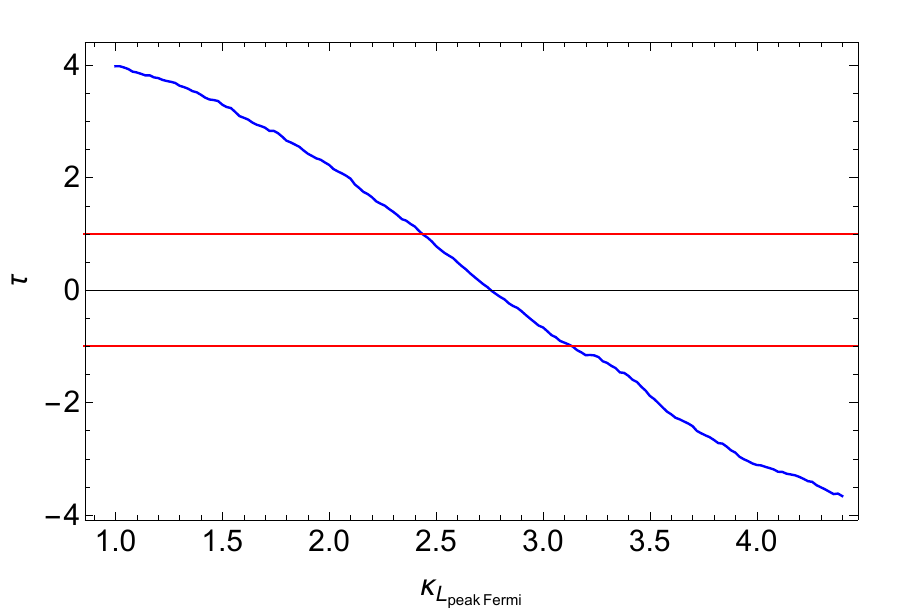}
    \caption{The \protect \cite{EffronPetrosian1992} $\tau$ statistic as a function of parameter $\kappa_{L_{\text{peak Fermi}}}$ (blue curve) used to de-transform the redshift evolution from $L_{\text{peak Fermi}}$-$z$ distribution. The red lines correspond to $\tau=\pm 1$, and establish a $\sim68.3\%$ confidence interval on the value of $\kappa_{L_{\text{peak Fermi}}}$ which removes the evolution.}
    \label{fig:Lpeaktau}
\end{figure}

One of the biggest challenges of contemporary astrophysics and cosmology is measuring the evolution of physical parameters characterizing astrophysical sources. The uniform description of astrophysical processes is a crucial step in standardizing the sources for cosmological measurements. The dependence of, e.g., the brightness of sources on the age of the Universe can introduce a significant bias, and thus, it has to be analyzed. The studies on evolution usually try to estimate the correlation between a given parameter and redshift. In the case of non-truncated data, such determination of correlation is a very straightforward procedure. However, astronomers almost always have to deal with nontrivial biases (e.g. flux-limited data). The flux-limited truncation is non-parallel to the axis of the luminosity-redshift diagram. Thus, it can very easily induce the correlation. One of the solutions is to test the correlation between volume and luminosity-limited subsets. This observation allowed authors of \cite{EffronPetrosian1992} to establish a selection-free test statistic inspired by Kendall's $\tau$:

\begin{equation}
    \tau = \frac{1}{\sqrt{N}}\sum^N_i\frac{R_i-E_i}{\sqrt{V_i}},
\end{equation}

where N is the total number of sources above the detection limit, $R_i$ is the rank of an i-th point (the number of sources in the associate set of i-th points with luminosity greater than the luminosity of i-th point). The associate set of an i-th source is the set: $\mathcal{A}_{i} = \{\mathbf{z, L}: z<z_{i} \And L>L_{min,\, i}\}$. Under the assumption that in case of no-correlation all the points are distributed uniformly, the expectation value of a rank is given by $E_i=\frac{1}{2}(n_i+1)$, and the variance $V_i = \frac{1}{12}(n^2_i-1)$, where $n_i$ is the number of elements in set $\mathcal{A}_i$.
This statistic has a normal distribution, where $\tau=0$ corresponds to a lack of correlation and $\tau \in \left[-1,1\right]$ is the $\sim 68.3\%$ confidence interval for the null hypothesis of a lack of correlation.
We assume the minimum observable flux by GBM is $F_{\text{lim}} = 2.5 \times 10^{-5} \, \rm{erg \, s^{-1} \, cm^{-2}}$, minimum energy fluence $f_{\text{lim, }Fermi} = 10^{-6}\, \rm{erg \, cm^{-2}}$, and the minimum observable peak energy $E_{\text{peak, lim}}=10^{1.6} \text{ keV}$.
We find $\tau \approx 5$ for raw $L_{\text{peak Fermi}}$ distribution, indicating that a null hypothesis of a lack of correlation between $L_{\text{peak Fermi}}$ and $z$ is rejected at $\sim 5 \sigma$ level.
Further, we employ $\tau$ as an estimator for the best-fit evolutionary function. We know that $\tau=0$ indicates no correlation (as in Kendall's $\tau$). Thus we search for such transformation of the distribution of $L_{\text{peak Fermi}}$ that gives $\tau=0$. We compute this statistic for the de-transformed variable $L'_{\text{peak, Fermi}} = \frac{L_{\text{peak, Fermi}}}{(1+z)^\kappa}$ over a grid of possible $\kappa$ values. This process allows us to obtain a selection-free description of the evolution, enabling us to accurately de-transform the redshift effect from our data. We demonstrate $\tau$ as a function of $\kappa$ in Fig. \ref{fig:Lpeaktau}. This result provides us with estimate $\kappa_{L_{\text{peak Fermi}}}=2.76^{+0.37}_{-0.32}$.

\bibliographystyle{elsarticle-harv} 
\bibliography{bibliography}






\end{document}